\documentclass[a4paper,fleqn,usenatbib]{mnras}

\usepackage{graphicx}	
\usepackage{amsmath}	
\usepackage{amssymb}	
\usepackage{gensymb}

\usepackage[switch,modulo]{lineno}

\title[Seismic modelling of $\theta$~Oph]{ Seismic modelling of early B-type pulsators observed by BRITE: I. $\theta$~Ophiuchi\thanks{Based on data collected by the BRITE Constellation satellite mission, designed, built, launched, operated and supported by the Austrian
Research Promotion Agency (FFG), the University of Vienna, the Technical University of Graz, the University of Innsbruck, the Canadian Space Agency (CSA), the University of Toronto Institute for Aerospace Studies (UTIAS), the Foundation for Polish Science \& Technology (FNiTP MNiSW), and National
Science Centre (NCN).}}

\author[Walczak et al.]{
Przemys{\l}aw Walczak$^{1}$\thanks{E-mail: walczak@astro.uni.wroc.pl},
 Jadwiga Daszy{\'n}ska-Daszkiewicz$^{1}$, 
 Andrzej Pigulski$^{1}$,
\newauthor Alexey Pamyatnykh$^{2}$,
 Anthony F.J. Moffat$^{3}$,
 Gerald Handler$^{2}$,
 Herbert Pablo$^{4}$,
\newauthor Adam Popowicz$^{5}$,
 Gregg Wade$^{6}$,
 Werner W. Weiss$^{7}$,
 Konstanze Zwintz$^{8}$\\
$^{1}$Astronomical Institute University of Wroc{\l}aw, Kopernika 11, 51-622 Wroc{\l}aw, Poland\\
$^{2}$Nicolaus Copernicus Astronomical Center Polish Academy of Sciences, Bartycka 18, 00-716 Warsaw, Poland\\
$^{3}$D\'ept. de physique, Univ. De Montr\'eal, C.P. 6128, Succ. Centre-Ville, and Centre de Recherche en Astrophysique du Qu\'ebec, \\ Montr\'eal, QC H3C 3J7, Canada\\
$^{4}$AAVSO, 49 Bay State Rd. Cambridge, MA 02138, USA\\
$^{5}$Institute of Automatic Control, Silesian University of Technology, Akademicka 16, Gliwice, Poland\\
$^{6}$Department of Physics \& Space Science, Royal Military College of Canada, PO Box 17000 Station Forces, Kingston, ON, Canada K7K 0C6\\
$^{7}$University of Vienna, Institute for Astrophysics, Tuerkenschanzstrasse 17, Vienna, Austria\\
$^{8}$Universit\"at Innsbruck, Institut f\"ur Astro- und Teilchenphysik, Technikerstrasse 25/8, A-6020 Innsbruck, Austria\\
}

\date{Accepted XXX. Received YYY; in original form ZZZ}

\pubyear{2015}

\begin{document}
\label{firstpage}
\pagerange{\pageref{firstpage}--\pageref{lastpage}}
\maketitle

\begin{abstract}
We analyse time-series observations from the BRITE-Constellation of the well known $\beta$ Cephei type star $\theta$~Ophiuchi.
Seven previously known frequencies were confirmed and nineteen new frequency peaks were detected.
In particular, high-order g modes, typical for the SPB (Slowly Pulsating B-type star) pulsators, are uncovered.
These low-frequency modes are also obtained from the 7-year SMEI light curve.\\
If g modes are associated with the primary component of $\theta$~Oph, then our discovery allows,  as in the case
of other hybrid pulsators, to infer more comprehensive information on the internal structure.
To this aim we perform in-depth seismic studies involving simultaneous fitting of mode frequencies, reproducing mode instability
and adjusting the relative amplitude of the bolometric flux variations.
To explain the mode instability in the observed frequency range a significant increase of the mean opacity in the vicinity of the $Z$-bump
is needed. Moreover, constraints on mass, overshooting from the convective core and rotation are derived.
If the low-frequency modes come from the speckle B5 companion then taking into account the effects of rotation
is enough to explain the pulsational mode instability.
\end{abstract}




\section{Introduction}

Seismic study of stars, called asteroseismology, is a unique way to find out which physical conditions prevail in deep stellar interiors.
This is done by  deriving constraints on input physics and stellar parameters used to compute evolutionary models.
One of the most important ingredients of stellar structure are opacities. These microphysics data still contain uncertainties, in particular in the vicinity of the metal ($Z$) bump, which was identified less than three decades ago \citep{1991ApJ...371L..73I,1992RMxAA..23...19S}.
This opacity maximum is caused mainly by intrashell transitions in the iron-group elements and occurs at depth corresponding to the temperature of $\log{T/\rm{K}}\approx5.3$.
In B-type main-sequence stars, the $Z$-bump activates the $\kappa$ mechanism that excites pulsations \citep{1992MNRAS.255P...1K,1992ApJ...393..272C,1992A&A...256L...5M}.
Thus, from detailed asteroseismic modelling of B-type pulsators one can expect to find clues for further improvements in the opacity data under conditions of stellar interiors.

The pulsational modelling is most challenging and potentially most rewarding for hybrid pulsators, i.e., stars which exhibit simultaneously low-order pressure (p)/gravity (g) modes as well as high-order gravity modes. In the case of early B-type stars, the lowest frequency g modes are not excited in standard-opacity models \citep[e.g.,][]{2004MNRAS.350.1022P,2017MNRAS.466.2284D}. As a possible remedy for this problem, the opacity increase in the Z-bump was proposed. The  example of pulsational studies with the modified opacity profile was provided by \cite{1996ApJ...471L.103C,1997ApJ...483L.123C} who tried to explain
oscillations in hot B-type subdwarfs (sdB stars). The opacity was modified by the enhancement of the iron abundance in the envelope. The physical mechanism which was proposed to be responsible for this enhancement, was radiative levitation. 
Another example of an  ad hoc iron enhancement in the $Z$-bump zone was done by \cite{2004MNRAS.350.1022P}, in order  to explain pulsations in the hybrid $\beta$\,Cep/SPB pulsator $\nu$\,Eridani.
In each case it has been found that opacity enhancement in the $Z$-bump would solve problems with mode excitation. A similar conclusion was obtained by \cite{2012MNRAS.422.3460S} who studied the excitation properties of $\beta$\,Cep stars in the Magellanic Clouds. The need for the opacity modification was obtained also for $\delta$ Scuti stars, that exhibit hybrid pulsations \citep{2015MNRAS.452.3073B}. To excite high-order g modes in these cooler pulsators, the opacity increase at the depth corresponding to $\log{T/\rm{K}}=5.06$ was needed. A new opacity bump at this temperature  was identified by \cite{2012A&A...547A..42C,2014A&A...565A..76C} in Kurucz model atmospheres \citep{Kurucz2004,2005MSAIS...8...14K,2011CaJPh..89..417K}.

Recently, \cite{2017MNRAS.466.2284D} interpreted the oscillation spectrum of $\nu$ Eri obtained from BRITE observations.
The innovative approach was applied, which involves simultaneous determination of seismic corrections to the model
and the mean opacity profile. As a result, a significant modification of the opacity profile around the $Z-$bump was obtained.
In particular, a huge increase at $\log{T/\rm{K}}=5.46$ (by 2 to 3 times) was indispensable to get instability of high-order g-modes. At this temperature nickel has its maximum contribution
to opacities at stellar conditions.
The number of possible opacity modifications was controlled by simultaneous fitting of the non-adiabatic parameter $f$, which describes the radiative flux variations at the level of the photosphere \citep{JDD2003,JDD2005}. A very good news was that there were very few models that met all three requirements, i.e., that adjust oscillation frequencies, reproduce instability of the observed frequency range and fit the empirical value of the parameter $f$.  Here, we apply the same approach to $\theta$~Ophiuchi that is very suitable for such asteroseismic analysis.
From the BRITE light curve we discovered a lot of g modes.  In the p-mode region we found one additional mode that allowed a firm identification of the mode degree, $\ell$, and azimuthal order, $m$, for all high frequency modes. This information gives hope that strong constraints on opacities, overshooting from the convective core and surface rotation can be obtained.

In Sect.\,\ref{sec:tOph} we introduce $\theta$~Ophiuchi. Sect.\,\ref{sec:observ} contains the description of observations, data reduction and frequency analysis.
In Sect.\,\ref{sec:sm} we present the results of our seismic modelling for both components of $\theta$~Oph. We end with conclusions in Sect.\,{\ref{sec:con}.

\section{$\theta$~Ophiuchi}
\label{sec:tOph}

$\theta$~Oph (HD\,157056, HR\,6453) is a bright ($V=3.248$ mag), hierarchical triple system.
The primary component, $\theta$~Oph A, contains a B2\,IV massive star ($\theta$~Oph Aa) and a low-mass ($M<1M_{\sun}$) spectroscopic companion ($\theta$~Oph Ab) with orbital period
of 56.71 d \citep{2005MNRAS.362..619B}. The third component, $\theta$~Oph B, was resolved in the triple system by means of speckle interferometry \citep{1993AJ....106.1639M,2002A+A...382...92S}. It is a B5 type star with a long orbital period of the order of 100 years \citep{2005MNRAS.362..619B}.

$\theta$~Oph A is a rather slow rotator with $V_{\rm{rot}}\sin{i}=31\pm3$ km\,s$^{-1}$ measured by \cite{1997A&A...319..811B}. This value was confirmed by \cite{2002ApJ...573..359A}
who derived $V_{\rm{rot}}\sin{i}=30\pm9$ km\,s$^{-1}$.
The metallicity determined from the low resolution ultraviolet spectra is $\text{[m/H]} = -0.15 \pm 0.12$ \citep{2005A&A...433..659N}, which corresponds to $Z=0.0095^{+0.0030}_{-0.0023}$. 
A slightly larger value of $Z$ was determined by \cite{2007MNRAS.381.1482B}, who found $Z=0.0114\pm0.0028$. 
Both determinations of $Z$, within the errors, are marginally consistent with solar metallicity, i.e. $Z_{\sun}=0.0134$ \citep{AGSS09}.

\begin{figure}
	\includegraphics[width=\columnwidth]{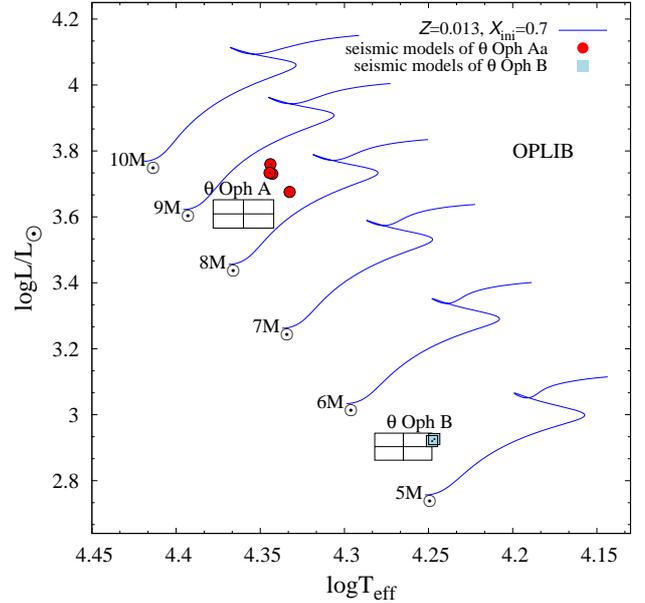}
	\caption{The HR diagram with the positions of $\theta$~Oph Aa and $\theta$~Oph B. Evolutionary tracks for masses from 5 to 10$M_{\sun}$ are depicted.
They were computed assuming OPLIB opacities, metallicity $Z=0.013$, initial hydrogen abundance $X_{\rm {ini}}=0.70$ and no overshooting from the convective core.
The remaining symbols will be explained in the text. 
}
	\label{fig:HRtOph}
\end{figure}

The star is a well known $\beta$ Cephei pulsator \citep{1922PDO.....8....1H,1956ApJ...124..168V,1957PASP...69..570M,1958ApJ...128..273V,2005ApJS..158..193S} and
was a target of extensive ground-based photometric \citep{2005MNRAS.362..612H} and spectroscopic \citep{2005MNRAS.362..619B} campaigns. Seven pulsational frequencies in the range \mbox{7\,--\,8 d$^{-1}$} were found in photometry. These frequencies are typical for $\beta$ Cep-type pulsations and correspond to the low-order p/g modes propagating in the stellar envelope. In the spectroscopic data only three modes were found. Successful mode identification \citep{2005MNRAS.362..612H} allowed for seismic modelling of the star. In the oscillation spectrum, the radial mode and all three components of the dipole mode were identified. Additionally, three out of five components of the $\ell=2$ mode were determined. The unambiguous identification of the azimuthal orders, $m$, of the $\ell=2$ components has been done from spectroscopy \citep{2005MNRAS.362..619B}.

The first attempt of seismic modelling of $\theta$~Oph has been carried out by \cite{2007MNRAS.381.1482B}. By fitting three centroid frequencies these authors constrained the basic parameters of the star, i.e., the mass, $M=8.2\pm0.3M_{\sun}$, central hydrogen abundance, $X_{\rm{c}}=0.38\pm0.02$, effective temperature, $T_{\rm{eff}}=22260\pm280$ K, rotational surface velocity $V_{\rm{rot}}=29\pm7$ km\,s$^{-1}$, and efficiency of the convective core overshooting, $\alpha_{\rm{ov}}=0.44\pm0.07$. Here, the overshooting parameter $\alpha_{\rm{ov}}$ is the size of the layer adjoined to the core and  affected by overshooting from the convective core. It is given in units of the local pressure scale height. These authors assumed metallicity $Z=0.0114$, derived from spectroscopic observations. It is worth to emphasize, that the centroid frequency of the quadrupole mode was not detected and its value, 7.2020 d$^{-1}$, was calculated by \cite{2007MNRAS.381.1482B} as the average of the frequencies of the two adjacent modes.

Seismic analysis carried out  by \cite{2009MNRAS.398.1961D} and \cite{2014IAUS..301..221W} confirmed the mode identification given by \cite{2005MNRAS.362..612H}. Moreover, the determination of the radial order, $n$, for the mode $\ell=0$, based on the empirical value of the parameter $f$ \citep{JDD2003,JDD2005}, became possible. The  mode is fundamental and this fact fixes the radial orders of the dipole and quadrupole modes: these are $p_1$ and $g_1$  modes, respectively.  
One of the most important results of this seismic analysis was a clear preference of the OPAL \citep{OPAL} opacity tables over the OP  \citep{OP} data. Models fitting the two well identified centroid frequencies and located within the observational error box of $\theta$~Oph A required the overshooting parameter  $\alpha_{\rm{ov}}\gtrsim0.3$.
\cite{2010A&A...515A..58L} performed two-dimensional calculations of the stellar evolution and pulsation  of $\theta$~Oph Aa. They derived a smaller value of the overshooting parameter, $\alpha_{\rm{ov}}=0.28\pm0.05$,
but the result was obtained with metallicity fixed  at $Z=0.02$.

In our studies we adopted the effective temperatures of $\theta$~Oph Aa, $\log{T_{\rm{eff}}}=4.360\pm0.018$, and $\theta$~Oph B, $\log{T_{\rm{eff}}}=4.265\pm 0.017$,  from \cite{2005MNRAS.362..612H}. Luminosities, $\log{L/L_{\sun}}=3.609\pm0.043$ for $\theta$~Oph Aa and $\log{L/L_{\sun}}=2.903\pm0.041$ for $\theta$~Oph B, were calculated using the Hipparcos parallax, $\pi=7.48\pm0.17$ mas \citep{2007A&A...474..653V}, bolometric correction from \cite{1996ApJ...469..355F} and visual magnitudes of $V=3.546$ mag ($\theta$~Oph Aa) and $V=4.876$ ($\theta$~Oph B) \citep{2002A+A...382...92S,2005MNRAS.362..612H}. We took into account also the colour excess $E(B-V)=0.017\pm0.008$ \citep{2005A&A...433..659N} to calculate the interstellar absorption. 
The corresponding error boxes of the components A and B of the $\theta$~Oph system are shown in the  Hertzsprung-Russell (HR) diagram in Fig.\,\ref{fig:HRtOph}. The evolutionary tracks were calculated with the Warsaw-New Jersey code \citep{1998A&A...333..141P} adopting the OPLIB opacities \citep{OPLIB2,OPLIB1}, metallicity $Z=0.013$ and initial hydrogen abundance $X_{\rm {ini}}=0.70$. The models were computed without taking into account the effects
of rotation and convective core overshooting. In all calculations the AGSS09 \citep{AGSS09} chemical mixture and the OPAL equation of state was used \citep[EOS\_2005,][]{2002ApJ...576.1064R}. We see, that both error boxes are on the main sequence. Moreover, the stars seem to be rather young. The masses derived from the HR diagram are around 5.5M$_{\sun}$ for $\theta$~Oph B and 8.5M$_{\sun}$ for $\theta$~Oph Aa.

\section{Observations and frequency analysis}
\label{sec:observ}
\subsection{BRITE photometry}\label{sec:brite}


The photometry analysed in the present paper has been obtained from space by the constellation of five BRITE (BRIght Target Explorer) nanosatellites \citep{BRITE1,2016PASP..128l5001P} during three runs in the Sagittarius field. In 2014 (the Sgr\,I field) only short test observations were carried out by a single BRITE satellite, the red-filter UniBRITE (UBr). They covered only about a month. During the two other seasons, 2016 (Sgr\,II), and 2017 (Sgr III), regular BRITE observations were performed lasting about six months each. These observations were secured by two red-filter, UBr and BRITE-Heweliusz (BHr), and two blue-filter, BRITE-Austria (BAb) and BRITE-Lem (BLb), BRITE satellites. Details of the BRITE observations are given in Table \ref{tab:brite-smei}. The Sgr\,I observations were obtained in staring mode, while the  Sgr\,II and Sgr\,III, in the chopping mode of observing \citep{2016PASP..128l5001P,2017A&A...605A..26P}. The images were analyzed by means of two pipelines described by \cite{2017A&A...605A..26P}. The resulting aperture photometry is subject of several instrumental effects \citep{2018adlc106} and needs pre-processing resulting in their removal. To remove instrumental effects we followed the procedure presented by \cite{2016A&A...588A..55P} with several modifications proposed by \cite{2018adlc175}. The whole procedure included rejection of outliers and the orbits with the largest scatter as well as one- and two-dimensional decorrelations with nine parameters of which eight were provided with the data (e.g.~position in the subraster or CCD temperature) while the ninth, satellite orbital phase, has been calculated. 
\begin{table*}
	\caption{Details of BRITE and SMEI data for $\theta$~Oph. $N_{\rm orig}$ and $N_{\rm final}$ stand for the original and final (after pre-processing) number of data points, respectively. RSD is the residual standard deviation after subtracting all significant modes. DT$_{(0-20)}$ is the detection threshold, calculated from the residual spectrum in the range (0, 20) d$^{-1}$. The number of modes is given in the preceding column.}
	\label{tab:brite-smei}
	\begin{tabular}{cccccrrcrr}
		\hline
		Field & Satellite &  Start & End & \multicolumn{1}{c}{Length of} & \multicolumn{1}{c}{$N_{\rm orig}$} & \multicolumn{1}{c}{$N_{\rm final}$} & \multicolumn{1}{c}{Modes}& \multicolumn{1}{c}{RSD} & \multicolumn{1}{c}{DT$_{(0-20)}$} \\
		&                  &     date   &  date   & \multicolumn{1}{c}{the run [d]} &&&  \multicolumn{1}{c}{detected}&\multicolumn{1}{c}{[mmag]} &\multicolumn{1}{c}{[mmag]}\\
		\hline
		Sgr I & UBr & 2014.04.29 & 2014.05.28 & 30 & 2952 & 2542  & 1 & 14.48 & 5.06 \\
		\hline
		Sgr II & BAb & 2016.04.21 & 2016.09.13 & 145 & 37532 & 30687 & 12& 10.49 & 0.69 \\ 
		& BHr & 2016.04.28 & 2016.09.20 & 146 & 86797 & 79027 & 25 & 6.02 & 0.27 \\
		\hline
		Sgr III & BAb & 2017.03.24 & 2017.09.16 & 176 & 46048 & 37020 & 12 & 12.95 & 0.75\\
		& UBr & 2017.04.02 & 2017.06.28 & 87 & 15028 & 12704 & 6 & 9.18 & 1.12 \\
		& BHr & 2017.08.05 & 2017.08.16 & 11 & 6518 & 5855 & 3 & 5.94 & 1.37\\
		\hline
		--- & SMEI & 2003.02.10 & 2010.11.26 & 2846 & 25274 & 18604 & 14 & 7.14 & 0.27 \\
		\hline
	\end{tabular}
\end{table*}

\subsection{SMEI photometry}\label{sec:SMEI}
$\theta$~Oph was also observed by the Solar Mass Ejection Imager (SMEI), the experiment placed on-board the Coriolis spacecraft \citep{2003SoPh..217..319E,2004SoPh..225..177J}. This experiment was designed to measure sunlight scattered by free electrons of the solar wind. A by-product of these observations was the photometry of bright stars, available through the University of California San Diego web page\footnote{http://smei.ucsd.edu/new\_smei/index.html}. The SMEI cameras observed without any filter, which means that their spectral response was defined by the sensitivity of the detector -- a wide range between 450 and 950~nm \citep{2003SoPh..217..319E}. The photometry was obtained as explained by \cite{2007SPIE.6689E..0CH}.

SMEI photometry suffers from many instrumental effects causing the frequent occurrence of instrumental red noise. However, its big advantage is about 8 years long coverage (2003 -- 2010) and a high duty cycle. This results in a very good frequency resolution and a sub-mmag detection level. The usefulness of SMEI data as a supplementary source of photometry for stars observed by BRITEs has been shown in several BRITE papers, e.g.~by \cite{2017A&A...603A..13K}. Some details of the SMEI photometry are also given in Table \ref{tab:brite-smei}. The pre-processing of the SMEI photometry was performed by including correction for the one-year long instrumental variability followed by the outlier removal and detrending. This was done by calculating averages in time intervals, spline interpolation between the averages and removal of the outliers by means of $\sigma$~clipping. These steps were carried out iteratively. During subsequent iterations the time interval was shortened (from 100 days down to 3 days) and the value of $\sigma$ for clipping varied from 5 at the beginning to 3.5 at the end of the process. Finally, the evidently instrumental periodic signals at 1 and 2~d$^{-1}$ were removed from the data. The whole procedure affects the frequencies below $\sim$0.3~d$^{-1}$, because when one does detrending from 1000 down to 3 days the signals in that range are not trustworthy. The advantage is a lower detection threshold at higher frequencies.

\subsection{Preliminary frequency analysis}
\label{prel-analysis}
The BRITE and SMEI photometry corrected for instrumental effects were the subject of the subsequent time-series analysis. At the beginning, each BRITE dataset listed in Table \ref{tab:brite-smei} was analyzed separately. Since the sets differ significantly in length, number of data points and quality, the number of detected periodic terms ranged from one in the shortest Sgr\,I BAb data set up to 25 in the Sgr\,II BHr data. The analysis of the BRITE data reveals a very rich frequency spectrum of $\theta$~Oph, consisting of both high and low-frequency terms (presumably p and g modes, respectively). This makes $\theta$~Oph another hybrid $\beta$~Cep/SPB pulsating star. In Fig.\,\ref{fig:tOph_FTR} show frequency spectra for data combined from all fields (cf. Table\,\ref{tab:brite-smei}). The top and middle panels correspond to BAb and BHr data, respectively. 



The frequency spectrum of the SMEI data (the bottom panel of Fig.\,\ref{fig:tOph_FTR}) confirmed the hybridity of $\theta$~Oph; in total, 14 terms were detected in the SMEI data. A detailed comparison of the frequencies detected in SMEI and BRITE data will be presented in Sect.~\ref{sec:fos}. Prior to this, however, we discuss the effects of binarity on the frequency spectra.


\begin{figure}
	\includegraphics[width=\columnwidth]{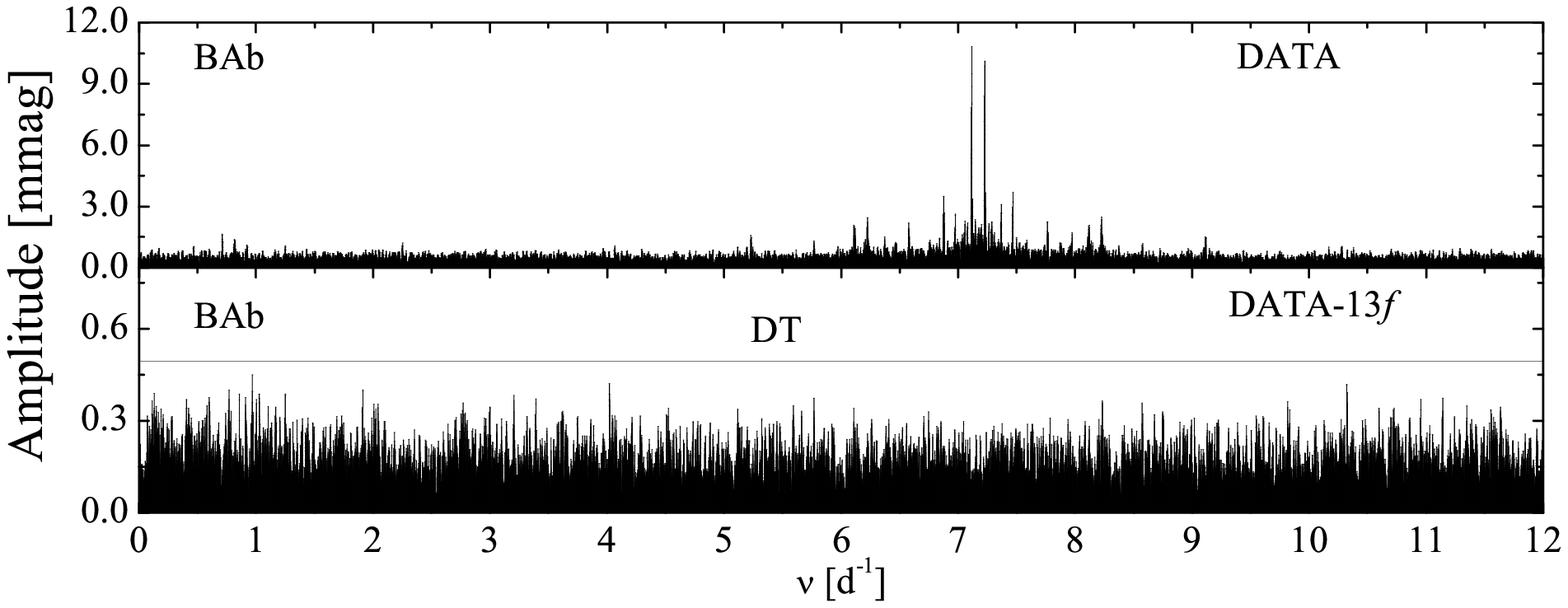}
	\includegraphics[width=\columnwidth]{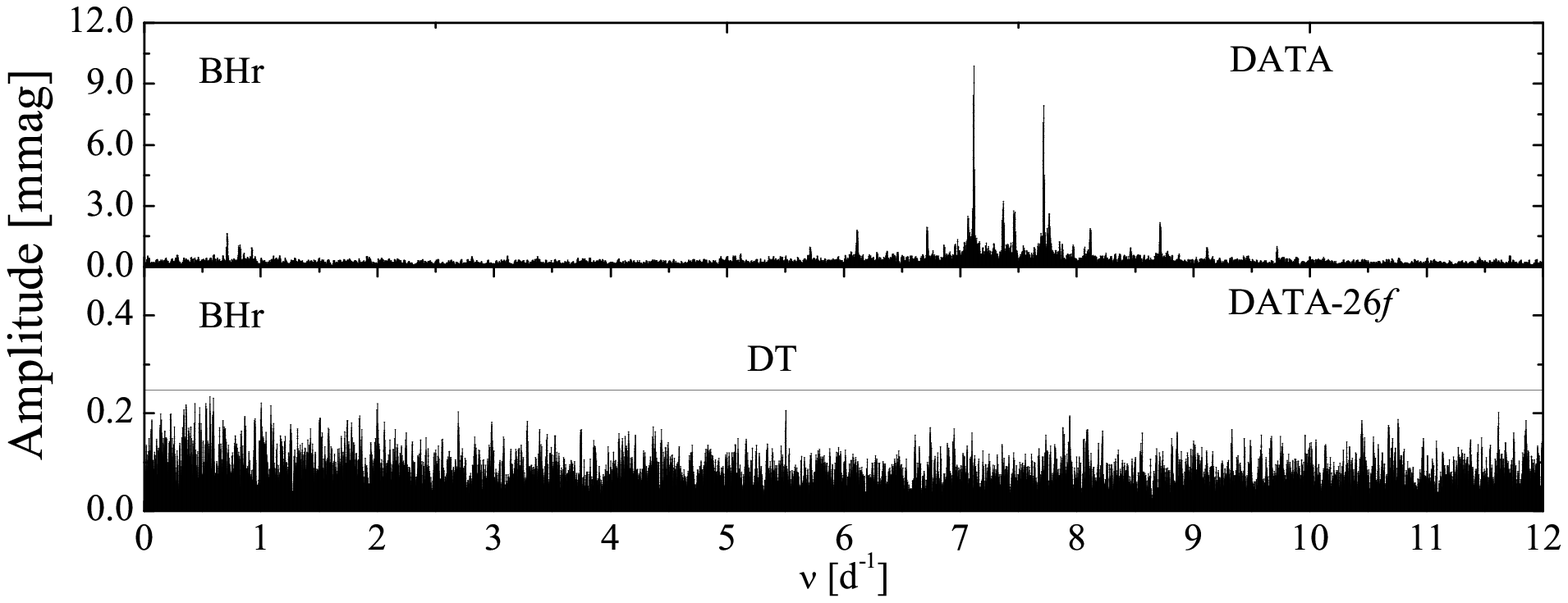}
	\includegraphics[width=\columnwidth]{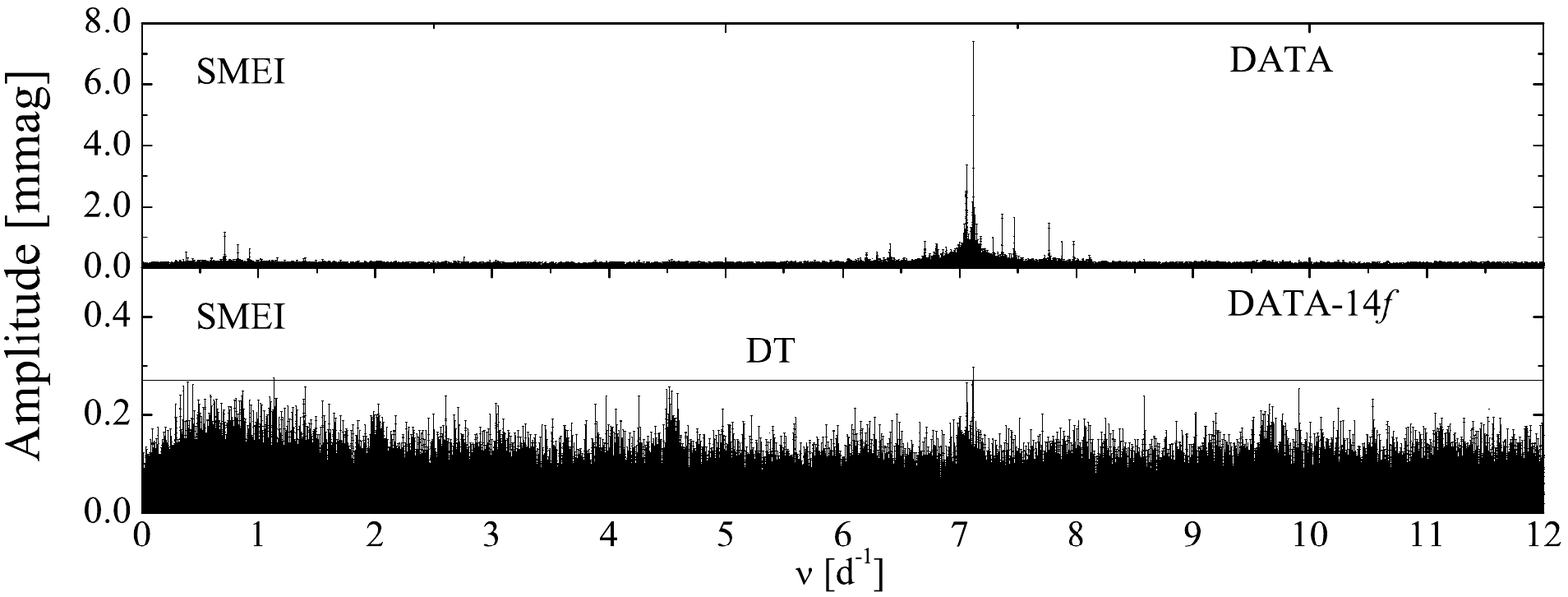}
	\caption{Top panels show the first run periodogram obtained from  all BAb data and  periodogram calculated from residuals after removing the variability with all significant frequencies. Middle and bottom panels --- the same for all BHr and SMEI data, respectively. Horizontal lines in each panel indicate the detection threshold.}
	\label{fig:tOph_FTR}
\end{figure}

 


\subsection{The light-time effect in the A\,--\,B system}
In systems which are wide enough for the light-time effect to become detectable, the effect can be used to discern which component pulsates. $\theta$~Oph is a hierarchical triple system. The component A is a single-lined spectroscopic (SB1) binary with a 56.7-day orbital period. Its spectroscopic orbital elements were derived by \cite{2005MNRAS.362..619B}. The light travel time through the projected range of the orbit of the Aa component around the centre of mass of the Aa\,--\,Ab system amounts to about 58\,s, i.e.~roughly 0.5\,per cent of the periods of the observed p modes and a much smaller fraction of the periods of g modes. This makes the light-time effect in this system practically undetectable with the present data. The situation is different for the wider A\,--\,B system. The system was first resolved by means of speckle interferometry by \cite{1993AJ....106.1639M}, who named it CHARA 182. Several other observations (see Sect.~\ref{sec:voab}) indicate a relatively short orbital period of the system, decades rather than centuries long. 

One may thus hope to obtain spectroscopic parameters of the A\,--\,B system from the study of the light-time effect. For this purpose, we performed analysis of the apparent variability of the pulsation periods detected in $\theta$~Oph Aa due to the light-time effect. In particular, we analyzed changes of the period of the dominant mode with frequency $\nu_1\approx$ 7.116~d$^{-1}$. The BRITE and SMEI data were divided into sufficiently long subsets to detect at least the three strongest modes in each of them. Then, for each subset, the times of maximum light were derived. The resulting O\,$-$\,C diagram is shown in Fig.\,\ref{fig:o-c}. It was calculated from the following ephemeris for the time of maximum light of the dominant mode 
\begin{equation}\label{eph1}
T_{\rm max,1} = \mbox{HJD}\,245000.0646 + 0.1405278 \times E,
\end{equation}
where $E$ is the number of pulsation cycles elapsed from the initial epoch. In addition, we made use of the photometric data of \cite{2005MNRAS.362..612H} and spectroscopic data of \cite{2005MNRAS.362..619B}. The spectroscopic data were of sufficient quality to derive the times of maximum radial velocity (RV) for the dominant p mode ($\nu_1$). This was done separately for seasons 2000\,--\,2001 and 2003. In order to combine photometric and radial-velocity times of maximum in the same O\,$-$\,C diagram, one has to know the phase lag between both curves. For $\beta$~Cep stars, light curve lags about 1/4 of the pulsation period behind the RV curve (the phase of maximum light corresponds to the phase of minimum radius and the mid-point of the decreasing branch of the RV curve). Fortunately, the 2003 RV data of \cite{2005MNRAS.362..619B} overlap with SMEI photometry. By trial and error, we derived the phase lag equal to 0.0340~d $\approx$ 0.242\,$P_1$, where $P_1 = 1/\nu_1$ is the pulsation period of the strongest mode. The spectroscopic elements were derived by fitting the equation
\begin{equation}\label{abtau}
(\mbox{O} - \mbox{C})(t) = A + B(t - t_{\rm mid}) + \tau(t,P_{\rm orb}, e, \omega, T_0, a_1\sin i),
\end{equation}
where $t$ is time, $t_{\rm mid} = \mbox{HJD}\,2453000$, arbitrarily chosen mid-time observation epoch, $A$ and $B$ are parameters allowing for a vertical shift and a linear trend in the O\,$-$\,C diagram, respectively. They can be interpreted as corrections to the initial epoch and period in the ephemeris (\ref{eph1}). The third factor in Eq.~(\ref{abtau}), $\tau$, is the light travel time, which depends on the orbital parameters: orbital period, $P_{\rm orb}$, eccentricity, $e$, longitude of periastron, $\omega$, time of periastron passage, $T_0$, semimajor axis of the orbit of the pulsating star around the centre of mass, $a_1$, and the inclination of the orbit, $i$ \citep{1952ApJ...116..211I}. The parameters obtained from the fit are given in the second column of Table \ref{tab:paroc}. They indicate a highly eccentric orbit of the $\theta$~Oph A\,--\,B system with a period of about 14 yr. While the spectroscopic parameters were derived from the times of maximum light and RV of the dominant mode, Fig.~\ref{fig:o-c} shows also the apparent changes of periods for two other p modes, $P_2 = 1/\nu_2 =$ 0.1339119~d, and $P_3 = 1/\nu_3 =$ 0.1356945~d. The changes of periods of all three modes are consistent, which is a very strong argument in favour of the light-time effect as the cause of these changes and the proof that --- as expected --- all three p modes originate in the same component ($\theta$~Oph Aa). In addition to orbital parameters Table \ref{tab:paroc} provides also the related values, half-range of the RV variability of the A component, $K_1$, and the mass function, $f(M)$. The value of $K_1$ indicates in particular, that RV changes due to the orbital motion in the A\,--\,B system should be detectable provided that sufficient RV data (especially around the periastron passage) will be obtained.
\begin{figure}
	\centering
	\includegraphics[width=\columnwidth]{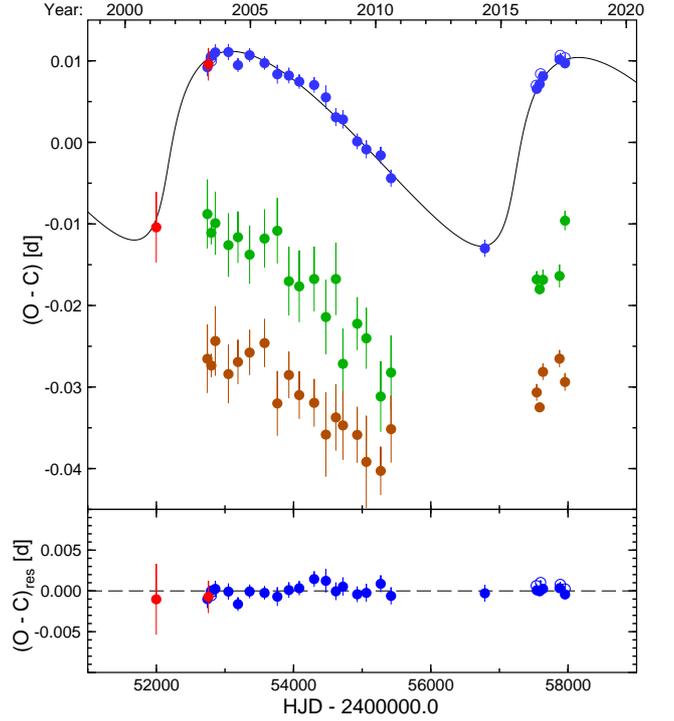}
	\caption{Top: The O\,$-$\,C diagram for the dominant mode ($\nu_1$) in $\theta$~Oph based on the photometric (blue dots) and RV (red dots) times of maximum. The O\,$-$\,C  values were calculated with respect to the ephemeris (\ref{eph1}). A constant phase shift of +0.0340~d was applied to the values of O\,$-$\,C derived from the RV data. The continuous line shows the fit of Eq.~(\ref{abtau}) to the O\,$-$\,C data. The open circles correspond to data in passbands other than BRITE red, SMEI and Str\"omgren $y$. They were not used in the fit. The green and brown dots represent O\,$-$\,C values for $P_2$ and $P_3$, respectively. They were shifted arbitrarily for a better visibility. Bottom: The residuals from the fit to P1.}
	\label{fig:o-c}
\end{figure}

\begin{table}
	\centering
	\caption{Parameters of spectroscopic and astrometric orbits of the $\theta$~Oph A\,--\,B (CHARA 182) system.}
	\label{tab:paroc}
	\begin{tabular}{cr@{\,$\pm$\,}lc}
		\hline
		Parameter & \multicolumn{3}{c}{Value}\\
		& \multicolumn{2}{c}{O\,$-$\,C data} & Vis data \\
		\hline
		$A$ [d] & $-$0.0005 & 0.0014 & --- \\
		$B$ & ($-1.5$& 1.7) $\times$ 10$^{-7}$ & ---\\
		$T_0$ [HJD] & 2457220 & 80& 2456790\\
		$\omega_{A,B}$ [\degr] & 348 & 6& 196\\
		$e$ & 0.72 & 0.16& 0.77\\
		$P_{\rm orb}$ [d] & 5070 & 170& 4160\\
		$P_{\rm orb}$ [yr] & 13.9 & 0.5& 11.4\\
		$a_1\sin i$ [R$_{\sun}$] & 610 & 170& ---\\
		$a$ [$^{\prime\prime}$] &\multicolumn{2}{c}{---}&0.28\\
		$\Omega$ [\degr] &\multicolumn{2}{c}{---}& 243\\
		$i$ [\degr] &\multicolumn{2}{c}{---}& 80\\
		\hline
		$K_1$ [km\,s$^{-1}$]&  8.8& 3.8& --- \\
		$f(M)$ [M$_{\sun}$] & 0.35 &0.45& ---\\
		\hline
	\end{tabular}
\end{table}

\begin{table*}
	\centering
	\caption{Observations of the relative positions of the A and B components of $\theta$~Oph.}
	\label{tab:vor}
	\begin{tabular}{lcccl}
		\hline
		Date & Separation & Polar angle & Band & Source, technique, remarks\\%
		~[yr] & [arcsec] & [\degr] & [nm] & \\
		\hline
		1991.25 & --- & --- & 511 & Hipparcos Catalogue \citep{1997ESASP1200.....E}, unresolved \\
		1992.4493 & 0.153 & 234.2 & 549 & \cite{1993AJ....106.1639M}, speckle interf.\\
		2000.40 & 0.300 $\pm$ 0.025 & 251.6 $\pm$ 3.3 & 2200 & \cite{2002A+A...382...92S}, adaptive optics\\
		2005.3567 & --- & --- & 880 &  \cite{2018MNRAS.473.4497R}, adaptive optics, unresolved\\
		2010.322 & $<$0.027 & {\sl V230.9} & & \cite{2012JDSO...8..313L}, lunar occ., vector angle is given in the third column\\  
		2010.322 & $<$0.028 & {\sl V234.9}& & \cite{2012JDSO...8..313L}, lunar occ., vector angle is given in the third column\\ 
		2014.2246 & 0.0623 & 61.2 & 692 & \cite{2015AJ....150..151H}, speckle interf., 180\degr added to the original polar angle\\
		2014.2246 & 0.0640 &  54.1 & 880 & \cite{2015AJ....150..151H}, speckle interf. \\
		2014.3036 & 0.0651 $\pm$ 0.0008 & 57.9 $\pm$ 0.6 & 543 & \cite{2015AJ....150...50T}, speckle interferometry\\
		\hline
	\end{tabular}
\end{table*}

\subsection{Astrometric orbit of the A\,--\,B system}\label{sec:voab}
Independently of the determination of the spectroscopic elements of the A\,--\,B system orbit from the O\,$-$\,C data, we checked if the available measurements of the relative positions of the A and B components are consistent with this orbit. The measurements were taken from the Fourth Catalog of Interferometric Measurements of Binary Stars\footnote{https://www.usno.navy.mil/USNO/astrometry/optical-IR-prod/wds/int4} and are presented in Table \ref{tab:vor}. As mentioned above, the system was first resolved by means of speckle interferometry by \cite{1993AJ....106.1639M} at a separation of about 0.15~arcsec. \cite{2002A+A...382...92S} measured a factor two times larger separation using adaptive optics in the infrared. Finally, two nearly simultaneous speckle observations by \cite{2015AJ....150..151H} and \cite{2015AJ....150...50T} resolved the system at a separation barely exceeding 0.06 arcsec. The observations span 22 years  (Table \ref{tab:vor}) and the changes of the relative positions indicate that the orbital period is much shorter than a century as suggested by \cite{2005MNRAS.362..619B}. On the other hand, the binary was not resolved by Hipparcos \citep{1997ESASP1200.....E}, lunar occultation in 2010 \citep{2012JDSO...8..313L} and speckle interferometry of \cite{2018MNRAS.473.4497R}. The five measurements made in four epochs allow one to derive the parameters of the visual orbit. The parameters of the astrometric orbit are presented in the third column of Table \ref{tab:paroc}. We do not provide their uncertainties as they cannot be reliably estimated due to the scarcity of the astrometric data. The astrometric orbit of $\theta$~Oph AB is shown in Fig.\,\ref{fig:vorb}. While the Hipparcos non-detection can be understood because of the small separation at the time Hipparcos observed the binary, the other two non-detections can be probably explained only by a relatively large magnitude difference between components A and B, close to the limits of detection.
\begin{figure}
	\centering
	\includegraphics[width=\columnwidth]{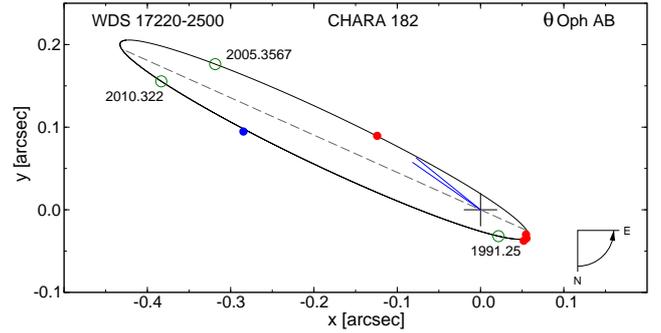}
	\caption{The apparent orbit of the $\theta$~Oph AB (CHARA 182) system. The ellipse corresponds to the solution given in Table \ref{tab:paroc}. The individual measurements are plotted with red (speckle interferometry) and blue (adaptive optics) dots. Open green circles mark the epochs of non-detections (labeled). 
		Large plus marks the position of component A, the dashed line is the line of apsides, the two blue lines are vector lines for lunar occultation observation \citep{2012JDSO...8..313L}. }
	\label{fig:vorb}
\end{figure}

The parameters of the astrometric orbit do not agree with those derived from the O\,$-$\,C data. In particular, the orbital period is $\sim$17\% shorter, there is also a $\sim$1.2~yr difference in the time of periastron passage and 28$\degr$ difference in the longitude of periastron (note that spectroscopic parameters include $\omega_A$, the parameter of the orbit of the A component around the centre of mass of the A\,--\,B system, while astrometric data provide $\omega_B = \omega_A - 180\degr$, the parameter of the secondary's relative orbit). Nonetheless, the two orbits presented in Table \ref{tab:paroc} can be regarded as fairly consistent given the fact that (i) the O\,$-$\,C data cover only slightly more than one orbital period, (ii) no obvious intrinsic period changes are present, (iii) the visual data are very scarce. We made attempts to fit simultaneously the O\,$-$\,C and astrometric data and we obtained some reasonable fits. However, the final result depends strongly on the relative weighting of the data. We therefore decided not to show solutions obtained from the combined data.

\subsection{The final frequency analysis}\label{sec:fos}
The light-time effect leads to the apparent changes of periods (Fig.\,\ref{fig:o-c}). For the modes originating in the A component the range of $\tau$ in Eq.~(\ref{abtau}), $\Delta\tau$, amounts to about 34 minutes, which corresponds to 0.17$P_1$. Although we will not combine SMEI and BRITE data\footnote{The main argument against combining SMEI and BRITE data is that low frequencies in the SMEI time-series are affected by the correction for instrumental effects while this is not the case for the BRITE data. The result of combining the data would be therefore unpredictable.}, it seems plausible to correct the observed times for this effect prior to the final frequency analysis. In principle, such a correction can be effectively applied only if all modes originate in the same component. We know from Fig.\,\ref{fig:o-c} that this is the case for the three strongest p modes. Following \cite{2005MNRAS.362..612H}, it is reasonable to assume that all p modes originate in the B2\,IV Aa component of $\theta$~Oph, because this is the only component of the system hot enough to excite these modes. On the other hand, g modes may originate in both the Aa and B components. The attempts to find in which component the g modes originate using the O\,$-$\,C diagram failed because their periods are much longer than $\Delta\tau$ and their amplitudes are too small. Nevertheless, there are two arguments that allow us to apply the $\tau$ correction to the SMEI and BRITE data despite the lack of knowledge in which component the g modes originate. First, it is much more likely that they originate in the Aa component because the contribution of the B component to the total light, estimated from the magnitude differences measured by means of speckle interferometry and adaptive optics imaging, amounts to 6\,--\,10\,\% in the visual and near infrared domains. Next, even if they originate in the B component, the correction applied as if they were excited in the Aa component is negligible in comparison with their periods. Therefore, the frequency spectra will not be affected.

In consequence, we corrected the observed times in the SMEI and BRITE data assuming that all variability originates in the Aa component. The corrected data were used in the final frequency analysis. We started the analysis with the BRITE data. In order to lower the detection threshold as much as possible, all BRITE data from three seasons were combined. Since amplitudes in the blue and red BRITE bands may differ, we followed the procedure used by \cite{2017MNRAS.464.2249H}. In short, the procedure consists of the following steps. The first frequency is detected in the combined blue and red data. The model consisting of a single sinusoidal term with this frequency is fitted separately to blue and red data. Then, the residuals are combined and used to search for the next frequency by means of the periodogram based on the discrete Fourier transform (DFT). The next sinusoid with the new frequency is added to the model and a full model consisting of all previously detected terms is fitted again to the blue and red data. The whole procedure is iterated until all significant terms are detected. The criterion we used to stop prewhitening is based on the signal-to-noise (S/N) ratio in the Fourier amplitude spectra, where S is the amplitude of the highest peak and N is the mean noise in the Fourier spectrum calculated in the range between 0 and 20~d$^{-1}$. We adopted S/N $=$ 4 as the plausible stopping criterion. We found 30 terms fulfilling this criterion in the combined blue- and red-filter BRITE data. The final model includes two additional combination frequencies detected with S/N slightly smaller than 4 because it is unlikely that they originate from the noise. The analysis of the corrected BRITE blue-filter, BRITE red-filter, and SMEI data was also carried out. As far as the number of detected terms is concerned, the results are only slightly different from the preliminary analysis (Sect.~\ref{prel-analysis}).

The results of the final frequency analyses are presented in Table\,\ref{tab:freq}.   The frequency changes due to the light-time effect were of the order of 0.0004~d$^{-1}$ for $p$ modes and of the order of 0.00001~d$^{-1}$ for $g$ modes. As expected, the largest number of terms was detected in the combined BRITE blue and red data. We detected eight modes in the high-frequency region, including all seven p modes ($\nu_1$ to $\nu_7$) found by \cite{2005MNRAS.362..612H}. The eighth mode is the fourth component of the quadrupole mode, $\nu_8$, which can be identified as $(\ell,m) = (2,0)$ mode, the centroid mode of the rotationally split quintuplet. Even more interesting is that $\theta$~Oph shows a very rich spectrum of g modes. In total, 20 independent low-frequency terms with frequencies between 0.11 and 2.77~d$^{-1}$ were found. Five of them were independently found in the BRITE and SMEI data. The sixth g mode detected in the SMEI data, $\nu_{\rm D}$, was not found in the BRITE data. The likely explanation is that the mode changed the amplitude. The amplitude change is probably responsible also for the non-detection of g modes in the frequency spectrum of the SMEI data despite the fact that they have relatively large amplitudes in the BRITE data; $\nu_{\rm L}$ is the best example.
\begin{table*}
	\caption{Oscillation frequencies and amplitudes derived from BRITE and SMEI data of $\theta$~Oph. The indexes for amplitudes, $A_i$, and S/N ratios, (S/N)$_i$, stand for: `cmb' -- combined blue- and red-filter BRITE data, `B' -- blue-filter BRITE data, `R' -- red-filter BRITE data, and `S' -- SMEI data.}
	\label{tab:freq}
	\begin{tabular}{clrrrrrrrl}
		\hline
		\multicolumn{2}{c}{Frequency} & \multicolumn{3}{c}{Amplitude [mmag]} & \multicolumn{4}{c}{Signal-to-noise ratio} & Notes\\ 
		ID & \multicolumn{1}{c}{[d$^{-1}$]} & \multicolumn{1}{c}{$A_{\rm {B}}$}& \multicolumn{1}{c}{$A_{\rm{R}}$} & \multicolumn{1}{c}{$A_{\rm{S}}$} & (S/N)$_{\rm cmb}$  & (S/N)$_{\rm B}$  &(S/N)$_{\rm R}$ & (S/N)$_{\rm S}$& \\
		\hline	
		$\nu_{\rm A}$ & 0.11308(10) & --- & 0.306 & --- & 5.46 & --- & 5.02 & --- & \\
		$\nu_{\rm B}$ & 0.24745(10) & --- & 0.303 & --- & 4.07 & --- & 5.10 & --- & \\
		$\nu_{\rm C}$ & 0.26792(13) & --- & 0.257 & --- & 4.04 & --- & 4.61 & --- & 1\,yr$^{-1}$\,alias higher in the BHr data \\
		$\nu_{\rm D}$ & 0.380505(24) & --- & ---  & 0.613 & --- & --- & --- & 9.07 & \\
		$\nu_{\rm E}$ & 0.40663(8) & --- & 0.386 & --- & 6.65 & --- & 6.14 & --- & \\
		$\nu_{\rm F}$ & 0.46650(11) & 0.756 & 0.693 & --- & 11.62 & 6.16 & 11.17 & --- & \\
		$\nu_{\rm G}$ & 0.56571(12) & --- & --- & --- & 4.63 & --- & --- & --- & \\
		$\nu_{\rm H}$ & 0.57247(10) & --- & 0.311 & --- & 5.81 & --- & 5.20 & --- & \\
		$\nu_{\rm I}$ & 0.60163(8)  & --- & 0.429 & --- & 7.80 & --- & 6.94 & --- & \\
		$\nu_{\rm J}$ & 0.711662(15) & 1.655 & 1.620 & 1.350 & 29.15 & 13.39 & 26.28 & 19.95 & the strongest g mode\\
		$\nu_{\rm K}$ & 0.76759(11) & --- & 0.296 & --- & 4.70 & --- & 4.81 & --- & \\
		$\nu_{\rm L}$  & 0.81071(7) & 1.165 & 0.982 & --- & 17.76 & 9.52 & 16.08 & --- & \\
		$\nu_{\rm M}$ & 0.823034(23) & 1.054 & 0.953 & 0.780 & 17.82 & 8.53 & 15.49 & 11.53 & \\
		$\nu_{\rm N}$ & 0.91490(10) & --- & 0.359 & --- & 5.58 & --- & 5.85 & --- & \\
		$\nu_{\rm O}$ & 0.92242(7) & 0.811 & 0.693 & 0.655 & 12.52 & 6.43 & 11.28 & 9.68 & \\
		$\nu_{\rm P}$ & 0.96999(10) & --- & 0.335 & --- & 5.44 & --- & 5.36 & --- & \\
		$\nu_{\rm Q}$ & 1.02923(9)  & --- & 0.386 & --- & 6.48 & --- & 6.04 & --- & \\
		$\nu_{\rm R}$ & 1.162098(33) & --- & 0.385 & 0.415 & 6.48 & --- & 6.13 & 6.14 & \\
		$\nu_{\rm S}$ & 1.91037(7) & --- & 0.442 & --- & 7.41 & --- & 7.13 & --- & \\
		$\nu_{\rm T}$ & 2.76640(18) & --- & 0.295 & 0.383 & 6.01 & --- & 5.43 & 5.66 & \\
		\hline
		$\nu_{1}$ & 7.1160281(28) & 10.146 &  9.591  & 7.149 &  181.73  & 81.96  & 154.31 & 105.66 & $(\ell,m) = (2,-1)$ \\
		$\nu_{8}$ & 7.20223(7)  & 0.499 & 0.395 & 0.294 & 8.20 & 4.05 & 6.54 & 4.35 &  $(\ell,m) = (2,0)$ \\
		$\nu_{5}$ & 7.28669(4)  & 1.152 & 1.230 & 0.941 & 22.48 & 9.26 & 19.82 & 13.90 &  $(\ell,m) = (2,+1)$ \\
		$\nu_{2}$ & 7.369490(10) & 2.697 & 2.550 & 1.930 & 42.41 & 21.93 & 41.21 & 28.52 &  $(\ell,m) = (2,+2)$ \\
		$\nu_{3}$ & 7.467599(19)  & 2.789 & 2.514 & 1.670 & 49.03 & 22.45 & 40.42 & 24.68 & radial mode\\
		$\nu_{4}$ & 7.765696(30) & 2.331 & 2.079 & 1.595 & 35.88 & 18.75 & 33.50 & 23.58 & $(\ell,m) = (1,-1)$\\
		$\nu_{6}$ & 7.874704(29) & 1.213 & 1.266 & 0.880 & 22.35 & 9.68 & 20.48 & 13.01 & $(\ell,m) = (1,0)$\\
		$\nu_{7}$ & 7.972558(33) & 1.016 & 1.141 & 0.921 & 20.12 & 8.31 & 18.36 & 13.61 & $(\ell,m) = (1,+1)$\\
		\hline
		$\nu_{7}-\nu_1$ & 0.856530 & --- & 0.547 & --- & 7.66 & --- & 8.81 & --- & \\
		$\nu_{4}+\nu_{\rm L}$ & 8.57641 & --- & 0.446 & --- & 7.50 & --- & 7.21 & --- &\\
		$\nu_{1}+\nu_{\rm M}$ & 7.93906 & --- & --- & --- & 3.75 & --- & --- & --- & \\
		$\nu_{1}+\nu_{\rm O}$ & 8.03845 & --- & 0.224 & --- & 3.88 & --- & 3.76 & --- & \\
		\hline
		$\sigma_A$ [mmag] & & 0.066& 0.030 & 0.074 & n/a & n/a & n/a & n/a &\\
		\hline
	\end{tabular}
\end{table*}

We also detected four combination terms. Three of them include one p and one g mode. This fact proves that at least the three g modes which form the combinations, $\nu_{\rm L}$, $\nu_{\rm M}$, and $\nu_{\rm O}$, are excited in the same component, $\theta$~Oph Aa.




\section{Seismic Models}
\label{sec:sm}


There are three well identified centroid frequencies in the oscillation spectrum of $\theta$~Oph Aa: the radial mode ($\nu_3$), a dipole mode ($\nu_{6}$) and a quadrupole mode ($\nu_{8}$). 
As we have mentioned in Sect.\,\ref{sec:tOph}, the radial mode is fundamental ($p_1$) as determined by \cite{2009MNRAS.398.1961D}. In such a case, $\nu_6$ can only be the $p_1$ mode, while $\nu_{8}$ the $g_1$ mode. Both $\nu_2$ and $\nu_{8}$ are low-order pressure modes propagating mainly in the outer envelope with small contribution from the gravity propagation zones near the core. 
The $m=0$ component of the $\ell=2,g_1$ mode, $\nu_8$, was detected for the first time. Its frequency is close the value 7.2020 d$^{-1}$ estimated by \cite{2007MNRAS.381.1482B}, although the difference is higher than the frequency error.

Our modelling consisted of calculating models fitting these three frequencies. For a given value of the metallicity and initial hydrogen abundance, this allowed us to constrain three free parameters: effective temperature, mass and overshooting parameter. In most of our models we assumed solar metallicity $Z=0.013$, which is, within the errors, consistent with the determinations of \cite{2005A&A...433..659N} and \cite{2007MNRAS.381.1482B}. We adopted also the standard initial hydrogen abundance $X_{\rm{ini}}=0.70$.

In principle, the observed pulsational frequencies can originate from either $\theta$~Oph Aa, $\theta$~Oph B or both components. The low mass component, $\theta$~Oph Ab, contributes a negligible amount of light to the total flux of the $\theta$~Oph A system. Therefore,  any light variations would have been undetectable and this possibility will not be discussed.  In the case of p modes, they are most probably associated with the primary component, $\theta$~Oph Aa. Firstly, the component B5 is of too late spectral type to have p modes excited. Next, given the magnitude difference measured by means of interferometry, it contributes only 5-10\% of the total flux in the visual domain. Finally, the splitting of p modes fits well to the observed
rotational velocity of $\theta$~Oph Aa (see Sect.\,\ref{sec:tOph_p_modes} for details). On the other hand, the origin of the majority of the g modes remains uncertain. Therefore, we will consider two scenarios. First, we will assume that all modes originate in $\theta$~Oph Aa (Sect.\,\ref{sec:tOphA}). Then, we will discuss the possibility of the  occurrence of g modes in $\theta$~Oph B (Sect.\,\ref{sec:tOphB}).

\subsection{The $\theta$~Oph Aa hypothesis}
\label{sec:tOphA}

\subsubsection{Standard opacity models}
\label{sec:tOph_std}
Evolutionary models were computed with the Warsaw-New Jersey code \citep{1998A&A...333..141P}. Then, pulsational calculations have been carried out with the customized non-adiabatic code of \cite{1977AcA....27...95D}.
Initially, we assumed the OPLIB opacity tables. The parameters of the models of $\theta$~Oph Aa that fit three centroid frequencies are given in Table\,\ref{tab:tOph:models}.
These models are marked as `$\theta$~Oph Aa standard' in the first column.

Depending on the assumed parameters, the model  mass of $\theta$~Oph Aa varies from about 8.0 to nearly $8.6M_{\sun}$.
The rotational velocity given in Table\,\ref{tab:tOph:models} represents the value of $V_{\rm {rot}}$ for which the splittings of the modes ($\ell=2,g_1$) and ($\ell=1,p_1$) are best reproduced (Sect.\,\ref{sec:tOph_p_modes}). We see, that the values of the order of 26-28 km\,s$^{-1}$ are appropriate for both detected multiplets. In Table\,\ref{tab:tOph:models} we give the parameters of seismic models for the three values of metallicity: $Z=0.010, 0.013$ and 0.020. The metallicity  and overshooting parameter are strongly correlated; higher $Z$ results in smaller $\alpha_{\rm {ov}}$.
A similar dependence was obtained by \cite{2007MNRAS.381.1482B}, \cite{2008MNRAS.385.2061D} and \cite{2009MNRAS.398.1961D}.
In the $\theta$~Oph Aa seismic models the parameter $\alpha_{\rm {ov}}$ decreases from about 0.39 for $Z=0.010$ to only 0.13 for $Z=0.020$. Higher metallicity results in slightly  higher mass and radius, and gives smaller effective temperature and luminosity. The effect of changing the initial hydrogen abundance, $X_{\rm{ini}}$, is quite significant as well. 
Larger hydrogen content implies smaller $\alpha_{\rm{ov}}$, $\log{T_{\rm {eff}}}$, $\log{L/L_{\sun}}$, and larger mass, radius and gravity. The value $X_{\rm {ini}}=0.725$ corresponds to the solar value \citep{2017A&A...607A..58B}, which should be similar for nearby early $B$-type stars \citep{2012A&A...539A.143N}. The model with higher hydrogen content is also older in comparison with a model calculated with the same metallicity and $X_{\rm {ini}}=0.70$.
All seismic models of $\theta$~Oph Aa are slightly outside the observational error box in the HR diagram (Fig.\,\ref{fig:HRtOph}).

In order to estimate the influence of the opacity data on the results of the calculations,  we constructed additional seismic models with the OP  and OPAL  tables. The model parameters are written in Table\,\ref{tab:tOph:models}. We see that the OPAL seismic model is similar to the OPLIB model calculated with the same value of $Z$ and $X_{\rm{ini}}$ with somewhat higher overshooting parameter. Much larger differences occur for the OP model, which has significantly lower mass, effective temperature, luminosity, radius, and gravity. It has also a higher value of $\alpha_{\rm {ov}}$ and is older. The results derived with the OPLIB and OPAL opacities seem to be fairly consistent. 

In the next step, we studied the instability properties of the OPLIB seismic models of $\theta$~Oph Aa.
The normalized  instability parameter, $\eta$, \citep{1978AJ.....83.1184S} is plotted as a function of the frequency in the upper panel of Fig.\,\ref{fig:tOph_nu_eta_std_A_B} for the standard opacity model with $Z=0.013$ and $X_{\rm {ini}}=0.70$. If the instability parameter is greater than zero the mode is unstable (excited) in a given model.  Modes with the degree $\ell$ up to 4 were considered. The vertical lines correspond to the observed frequencies of $\theta$~Oph Aa. Their height indicates the amplitude in the BRITE R filter (the scale is given on the right-hand Y axis).

As one can see, the frequencies $\nu_3=7.4676$ d$^{-1}$, $\nu_6=7.8746$  d$^{-1}$ and $\nu_{8}=7.2022$  d$^{-1}$ are well reproduced by this model. Also, the instability parameter in the high frequency region is positive indicating that the modes are unstable. The problem with the stability occurs for the low-frequency modes. In these models, nearly all modes with frequencies below about 4 d$^{-1}$ are stable. The pulsations of B-type stars are driven through the $\kappa$ mechanism \citep{1962ZA.....54..114B,1963ARA&A...1..367Z} operating in the $Z-$bump \citep{1993MNRAS.262..204D,1993MNRAS.265..588D}. The value of the opacity coefficient in the $Z-$bump, and hence the excitation efficiency, is very sensitive to the heavy element contents. The higher the $Z$ value is, the larger the instability. However, even for $Z=0.02$ the dipole and quadrupole modes in the g-mode region are stable. Higher degree modes become unstable but for frequencies larger than the observed values.

We would like to emphasize that the stellar model with about $8M_{\sun}$ calculated with standard opacity tables can have unstable g modes with $\ell=1$ and $\ell=2$, but for far more evolved stages, i.e., near the Terminal Age Main Sequence \citep{2017MNRAS.469...13S}. In the case of $\theta$~Oph Aa, the stellar parameters and the frequency range of p modes clearly indicate an early stage of evolution.

\begin{table*}
	\caption{The parameters of seismic models of $\theta$~Oph Aa and $\theta$~Oph B. In the first column there is a model name and the the following columns contain: metallicity $Z$, initial hydrogen abundance $X_{\rm {ini}}$, overshooting parameter $\alpha_{\rm {ov}}$, mass in solar units $M/M_{\sun}$, effective temperature $\log{T_{\rm {eff}}}$, luminosity $\log{L/L_{\sun}}$, radius in solar units $R/R_{\sun}$, surface gravity $\log{g}$, age $\log{t/\rm{yr}}$ (with $t$ given in years from the Zero Age Main Sequence), rotational velocity $V_{\rm {rot}}$ [km\,s$^{-1}$], and core hydrogen content $X_{\rm {c}}$.}
	\label{tab:tOph:models}
	\begin{tabular}{cccccccccccccc}
		\hline
		Model&$Z$  &  $X_{\rm {ini}}$ & $\alpha_{\rm{ov}}$  &  $M/M_{\sun}$ & $\log{T_{\rm{eff}}}$ & $\log{L/L_{\sun}}$ &  $R/R_{\sun}$ &    $\log{g}$  & $\log{t/\rm{yr}}$ & $V_{\rm {rot}}$ & $X_{\rm {c}}$\\\hline
		OPLIB:\\
		$\theta$~Oph Aa standard & 0.013 & 0.700 & 0.2957 & 8.396 & 4.3428 & 3.7304 & 5.053 & 3.9539 & 7.305 & 26.85 & 0.3924 \\ 
		$\theta$~Oph Aa standard & 0.013 & 0.650 & 0.3190 & 8.064 & 4.3542 & 3.7643 & 4.986 & 3.9480 & 7.253 & 26.54 & 0.3541 \\
        $\theta$~Oph Aa standard & 0.010 & 0.700 & 0.3900 & 8.216 & 4.3491 & 3.7500 & 5.021 & 3.9502 & 7.346 & 26.52 & 0.3743 \\
        $\theta$~Oph Aa standard & 0.020 & 0.700 & 0.1353 & 8.598 & 4.3279 & 3.6767 & 5.090 & 3.9580 & 7.241 & 27.15 & 0.4220 \\
        $\theta$~Oph Aa standard & 0.013 & 0.725 & 0.2843 & 8.565 & 4.3374 & 3.7144 & 5.087 & 3.9569 & 7.330 & 26.98 & 0.4117 \\
		\\
		
		$\theta$~Oph Aa modified & 0.013 & 0.700 & 0.2353 & 8.639 & 4.3440 & 3.7597 & 5.199 & 3.9416 & 7.264 & 27.78 & 0.3930 \\
		$\theta$~Oph Aa modified & 0.013 & 0.650 & 0.2608 & 8.305 & 4.3552 & 3.7944 & 5.137 & 3.9348 & 7.212 & 27.53 & 0.3552 \\
		$\theta$~Oph Aa modified & 0.010 & 0.700 & 0.3243 & 8.476 & 4.3506 & 3.7811 & 5.167 & 3.9386 & 7.304 & 27.48 & 0.3747 \\
		$\theta$~Oph Aa modified & 0.020 & 0.700 & 0.0726 & 8.813 & 4.3284 & 3.7029 & 5.233 & 3.9446 & 7.199 & 28.05 & 0.4217 \\
		$\theta$~Oph Aa modified & 0.013 & 0.725 & 0.2260 & 8.809 & 4.3386 & 3.7435 & 5.2305 & 3.9448 & 7.290 & 27.91 & 0.4121 \\

		\\
		$\theta$~Oph B standard & 0.013 & 0.700  &     0.0000 &     5.400  &  4.2466  &  2.9261 &   3.118 & 4.1642   & 7.305    &  150 & 0.5780 \\
		$\theta$~Oph B standard & 0.013  &   0.700 &   0.0000   &   5.400    &4.2480  &   2.9201  &  3.078   & 4.1756 &   7.264  &    150 &  0.5905 \\

		$\theta$~Oph B standard & 0.013 & 0.725 & 0.0000 & 5.400 & 4.2374 & 2.8735 & 3.062 & 4.1802 & 7.305 & 150 & 0.6156 \\
		$\theta$~Oph B standard & 0.013 & 0.725 & 0.0000 & 5.400 & 4.2386 & 2.8683 & 3.027 & 4.1902 & 7.264 & 150 & 0.6266 \\
	
		\hline		
		OPAL:\\
		$\theta$~Oph Aa standard&0.013 & 0.70 & 0.3121  &  8.413 &   4.3444  &  3.7334 &   5.034  &  3.9581  &  7.297 &   27.72 &   0.3970
		\\\hline
		OP:\\
		$\theta$~Oph Aa standard & 0.013 & 0.70 &  0.3870 &   8.001 &   4.3325  &  3.6756 &   4.975  &  3.9470 &   7.381 &   27.22  & 0.3843 \\ 
		\hline
	\end{tabular}
\end{table*}

\begin{figure}
    \includegraphics[width=\columnwidth]{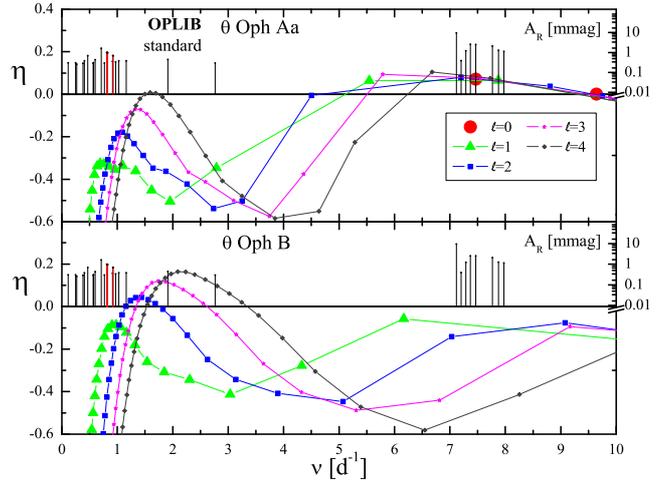}
    \caption{The instability parameter, $\eta$, as a function of the frequency for seismic models of $\theta$~Oph Aa (the upper panel) and $\theta$~Oph B (the bottom panel). Both models have $Z=0.013$ and $X_{\rm {ini}}=0.70$. Vertical lines indicate the observed oscillation spectrum of $\theta$~Oph. Their heights correspond to the amplitude in the BRITE R filter. Three g modes, $\nu_{\rm {L}}$, $\nu_{\rm {M}}$, $\nu_{\rm{O}}$, that certainly belong to the $\theta$~Oph Aa component were marked with red lines. The parameters of both models are given in Table\,\ref{tab:tOph:models}.}
    \label{fig:tOph_nu_eta_std_A_B}
\end{figure}

\subsubsection{Enhanced opacity models}
\label{sec:tOph_eom}

The purpose of our further seismic modelling was to search for appropriate modifications of the mean opacity profile in order to get instability in the whole range of the observed frequencies, in particular in the low-frequency range. Here, we follow the approach  of \cite{2017MNRAS.466.2284D} used in the analysis of the other hybrid pulsator, $\nu$ Eridani.
In this approach we increase or decrease the mean opacities at certain temperatures by adding a sum of Gaussian functions. The modified opacity profile had the following form:
$$\kappa (T)=\kappa_0(T) \left[1+\sum_{i=1}^N b_i \exp\left( -\frac{(\log T-\log T_{0,i})^2}{a_i^2}\right) \right],$$
where $\kappa_0(T)$ is the standard opacity profile and $(a,~b,~T_{0})$ are parameters of the Gaussian function. The parameters $a$ and $b$ describe the width and height, respectively, and $T_{0}$ the central temperature at which the opacity is changed. $N$ stands for the number of added opacity bumps. We search for an opacity modification in an extensive grid of stellar models. We used the following ranges of the free parameters: $(5.0,5.5)$ for $\log{T_0}$, $(0,1)$ for $a$ and ($-1,3)$ for $b$. In the initial calculations we used a step 0.01 in the $a$ parameter, 0.1 in the $b$ parameter and 0.01 in $\log{T_0}$. Then, the most interesting models were fine-tuned by means of grid calculations with smaller steps, i.e. 0.001 for $a$, 0.01 for $b$ and 0.001 for $\log{T_0}$.

We have three centroid frequencies in the p mode region that can be used in modelling. In principle, it is possible to find plenty of opacity-modified seismic  models which match these frequencies. \cite{2017MNRAS.466.2284D} have shown that adding the non-adiabatic parameter $f$ limits very effectively the number  of solutions. The parameter $f$ is defined as the ratio of the relative  bolometric flux perturbation to the relative radial displacement. The method of determination of the empirical value of the $f$-parameter is described by \cite{JDD2003,JDD2005}. It requires nearly simultaneous photometric and spectroscopic data. We applied this method to the observations gathered during campaigns dedicated to $\theta$~Oph \citep{2005MNRAS.362..612H,2005MNRAS.362..619B}. These data allowed us to determine the empirical value of $f$ for the radial mode.
Thus, our seismic modelling, which we called complex asteroseismology \citep{2009MNRAS.398.1961D}, consisted of  fitting the three centroid frequencies, getting the mode instability in the observed frequency range and reproducing the empirical value of $f$ for the radial mode.
As a result we obtain very few seismic models that meet the above conditions. One of the best solutions has the following opacity modification:
\begin{itemize}
   \item $\log{T_{0,1}}=5.060$, $a_1=0.082$, $b_1=0.30$,
   \item $\log{T_{0,2}}=5.300$, $a_2=0.500$, $b_2=0.65$,
   \item $\log{T_{0,3}}=5.470$, $a_3=0.077$, $b_3=1.45$.
\end{itemize}
The opacity increase at $\log{(T/\rm {K})}=5.470$ is mainly necessary to excite high-order g modes, whereas   the $\kappa$ increase at $\log{(T/\rm {K})}=5.300$ maintains the p-mode instability. The opacity modification at  $\log{(T/\rm {K})}=5.060$ is indispensable for fitting the parameter $f$. 

The models with the modified opacity  have a different structure. As a consequence, the opacity modified models that fit the three centroid frequencies have different parameters from the standard opacity models. The new models are marked in Table\,\ref{tab:tOph:models} as '$\theta$~Oph Aa modified'. In comparison with standard models, they have less effective overshooting from the convective core, higher masses and are of lower age.


The value of $f$ depends slightly on the atmosphere models. We used two sets of data: the LTE model atmosphere by \cite{Kurucz2004,2005MSAIS...8...14K,2011CaJPh..89..417K} and the non-LTE models by \cite{2007ApJS..169...83L}. We derived the following empirical values for the radial mode ($\nu_3$): $f=(f_{\rm{R}},f_{\rm{I}})=(-8.35\pm0.80, -0.11\pm0.82$) (LTE) and $f=(-8.08\pm1.07,0.12\pm1.09)$ (non-LTE). The parameter $f$ is complex for non-adiabatic pulsations, and the subscript R denotes the real part of $f$ while the subscript I its imaginary part. The theoretical value of the opacity modified seismic model for $Z=0.013$ and $X_{\rm {ini}}=0.70$ is $f=(-9.05, 0.68)$. We see, that this model fits the empirical value of $f$ regardless of the adopted model atmospheres. Other opacity modified seismic models of $\theta$~Oph Aa listed in Table\,\ref{tab:tOph:models} fit also the empirical values of the parameter $f$. The only exception is a model with $Z=0.010$, whose the real part of $f$ is too small in comparison with its empirical counterpart. The parameter $f$ was also reproduced by standard models (excluding the model with $Z=0.01$, whose $f_{\rm {R}}$ was too small). Thus, the modification of the opacity coefficient increased the excitation  and --- at the same time --- did not spoil the agreement between empirical and theoretical values of the parameter $f$. That is why the final profile required rather complicated changes.

The instability parameter, $\eta$, of the opacity-modified seismic model of $\theta$~Oph Aa, calculated with $X_{\rm {ini}}=0.70$ and $Z=0.013$, is shown in Fig.\,\ref{fig:tOph_om}.
We see that the instability occurs in both g and p mode regions. The detailed analysis of the instability in the vicinity of low frequencies is given later in Sect.\,\ref{sec:tOph_TA_gmodes}. For p modes, the instability parameter decreased  at higher frequencies but the observed modes are still within the theoretically excited region. Such a cut off of the instability at a frequency of about 8.5 d$^{-1}$ fits even better to the observed spectrum. The problem remains with $\nu_{T}\approx2.77$ d$^{-1}$, which can be explained only if we assume higher degree modes, $\ell\gtrsim6$.

The required modification of the opacity profile is quite complicated, as can be seen in Fig.\,\ref{fig:tOph_kappa_om}, where we show a comparison of the standard and modified mean opacity coefficient ($\kappa$) and its temperature derivative ($\kappa_T=\partial\log{\kappa}/\partial \log{T}$). In this seismic model we have a 30\% increase of the opacity coefficient near  $\log{T_{0,1}}=5.060$,
65\% at $\log{T_{0,2}}=5.300$ and the largest increase, 145\% at $\log{T_{0,3}}=5.460$. This last value, at the so-called 'nickel' opacity, was enforced by the condition on the instability
of high-order g modes. The opacity changes have significant influence on the temperature derivative of the $\kappa$ coefficient, which is directly related to the excitation properties of pulsational modes. There is a sudden drop in $\kappa_{\rm {T}}$, near $\log{T/\rm{K}}\approx5.5$,  followed by a large increase. Such a structure works very effectively in driving  stellar  pulsations \citep{1999AcA....49..119P}.

\begin{figure}
    \includegraphics[width=\columnwidth]{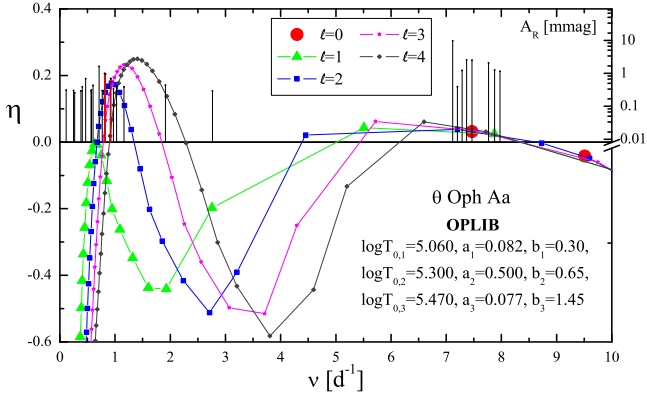}
     \caption{The instability parameter, $\eta$, as a function of the frequency for the opacity modified seismic model of $\theta$~Oph Aa calculated with $X_{\rm {ini}}=0.70$ and $Z=0.013$. Models with $\ell$ up to 4 were included. The opacity modification is given in the legend while the model parameters are given in Table\,\ref{tab:tOph:models}. }
    \label{fig:tOph_om}
\end{figure}

\begin{figure}
	\includegraphics[width=\columnwidth]{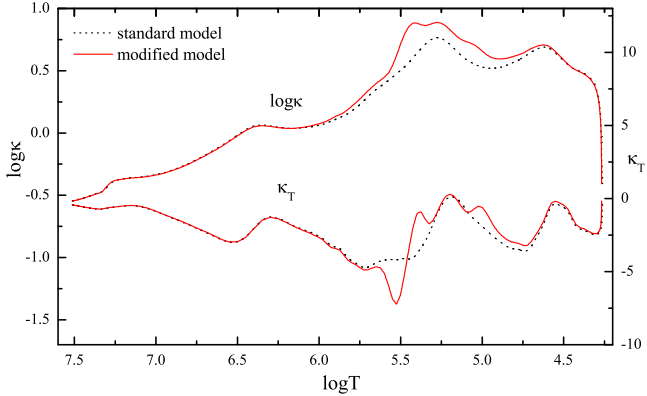}
    \caption{Comparison of the mean opacity profile ($\kappa$) and its temperature derivative ($\kappa_T$) for the standard and modified opacity models of $\theta$~Oph Aa calculated with $X_{\rm {ini}}=0.70$ and $Z=0.013$. The model parameters are given in Table\,\ref{tab:tOph:models}, the opacity modification is described in the text.}
    \label{fig:tOph_kappa_om}
\end{figure}

\subsubsection{Rotation effects on g modes}
\label{sec:tOph_TA_gmodes}

The primary component of the $\theta$~Oph system rotates rather slowly. Nevertheless, it is worth to check the effects of rotation on g modes since even slow rotation can affect their properties if pulsational frequencies are comparable to the rotational frequency \citep{2003MNRAS.343..125T}. Here we apply the Traditional Approximation \cite[TA,][]{1970attg.book.....C,1989nos..book.....U,1996ApJ...460..827B,1997ApJ...491..839L,2003MNRAS.340.1020T,2003MNRAS.343..125T,2007MNRAS.374..248D,2007AcA....57...11D} to include the effects of the Coriolis force.
In the case of g modes, we do not have constraints on the degree $\ell$. Unfortunately, the method of mode identification based on the BRITE amplitudes and phases did not give unique constraints on the mode degree $\ell$. Therefore, we focus only on their instability properties.

We used the recently developed non-adiabatic pulsational code (Walczak et al., in preparation). The code includes the effects of rotation in the framework of TA and is fully non-adiabatic throughout the whole star. In this respect, it is different from  the Dziembowski's code, which is quasi-adiabatic in the deep stellar interior and non-adiabatic in the outer layers. 
The non-adiabatic effects are more important for high-order g modes because the longer pulsation periods allow for more efficient heat exchange. The instability parameter for high-order g modes is slightly smaller. The new code, however, is much slower than Dziembowski's code so it was used only for a calculation of the limited number of most interesting models. The calculation of a model grid with the new code was not possible due to limitation of computational resources.
In the case of low-order p and g modes both codes produce nearly the same results.

In Fig.\,\ref{fig:tOph_modif_TA}, we show the instability parameter for the modified seismic model calculated with $X_{\rm {ini}}=0.70$, $Z=0.013$ and for the modes with $\ell=1$ and 2.
The value of the rotational velocity amounts to 28 km\,s$^{-1}$, which reproduce the rotational splitting of the dipole and quadrupole p modes. Other model parameters are given in Table\,\ref{tab:tOph:models} ($\theta$~Oph Aa modified). For a better clarity we did not depict modes with $\ell > 2$. The azimuthal order, $m$, changes from $-\ell$ to $+\ell$. In our convention, the retrograde modes have negative values of $m$. 

\begin{figure}
	\includegraphics[width=\columnwidth]{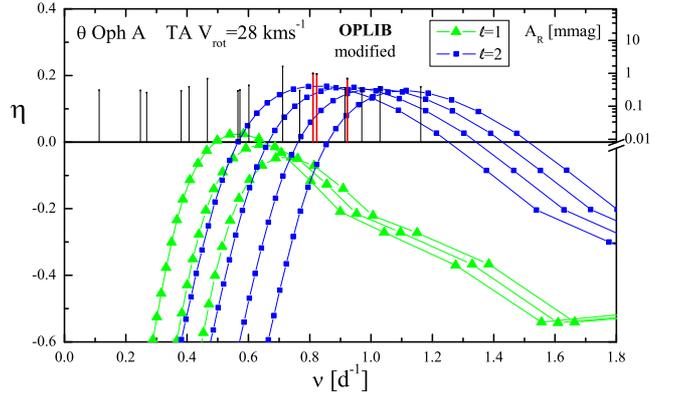}
	\caption{The instability parameter, $\eta$, as a function of the frequency for the seismic model with modified opacity calculated with $X_{\rm {ini}}=0.70$ and $Z=0.013$. Only the g-mode region is shown. The effects of  rotation were included by means of the traditional approximation assuming a rigid rotation with $V_{\rm{rot}}=28$ km\,s$^{-1}$.}
	\label{fig:tOph_modif_TA}
\end{figure}

The maximum value of $\eta$ for the dipole modes depends on the azimuthal order, $m$. The largest instability occurs for the retrograde modes ($m=-1$) and for frequencies from about 0.4 d$^{-1}$ to about 0.6 d$^{-1}$ the modes are excited. All axisymmetric and prograde dipolar modes are stable in this model. The quadrupole modes are unstable in a large frequency range and
the maximum of $\eta(\nu)$ is actually independent of $m$. As we can see, this model does not explain the six observed modes with the lowest frequencies.


There are frequencies in the g mode domain that may form  quasi-equally spaced sequence in periods. The modes involved in the sequence are: $\nu_{R}\approx1.162$ d$^{-1}$ (0.859 d), $\nu_{Q}\approx1.028$ d$^{-1}$ (0.973 d), $\nu_{N}\approx0.915$ d$^{-1}$ (1.095 d), $\nu_{M}\approx0.823$ d$^{-1}$ (1.215 d), $\nu_{K}\approx0.768$ d$^{-1}$ (1.302 d) and $\nu_{J}\approx0.712$ d$^{-1}$ (1.405 d).

Unfortunately, the sequence, if it does consist of modes with the same degree and consecutive radial orders, cannot be unambiguously identified. For example, it can be fitted with modes $(\ell,m)$ = $(2,+1)$, $(2,-2)$,  $(3,-1)$, $(3, -2)$ or $(3,-3)$.  The modes could be used in detailed seismic modeling provided that successful and independent mode identification is possible.

\subsubsection{Rotational splitting of p modes}
\label{sec:tOph_p_modes}

The oscillation spectrum of $\theta$~Oph Aa contains two well identified  multiplets. There are four components of the $\ell=2$ quintuplet and all components of the $\ell=1$ triplet. Splittings in quadrupole and dipole modes are different, however. For the $\ell=2$ modes, the splitting is of the order of 0.08 d$^{-1}$, while for the $\ell=1$ modes it amounts to about 0.10 d$^{-1}$.  It turns out that these splittings can be explained by one value of the rotational rate. In the first order approximation, the value of the frequency of the non-axisymmetric modes is given by:
$$
\nu_{\ell,n,m}=\nu_{\ell,n}+(1-C_{n\ell})\nu_{\rm {rot}},
$$
where $\nu_{\ell,n}$ is the frequency of the centroid mode, $C_{n\ell}$ is the Ledoux constant \citep{1951ApJ...114..373L} and $\nu_{\rm {rot}}$ is the rotational frequency.

The Ledoux constant in the models of $\theta$~Oph Aa differs significantly for $\ell=1$ and $\ell=2$ modes.
The value of $C_{n\ell}$ amounts to about 0.035 for the $\ell=1, p_1$ mode and 0.208 for the $\ell=2,g_1$ mode.
This implies larger splitting for the dipole modes. In Fig.\,\ref{fig:tOph_nu_eta_Cnl}, we show the observed frequencies
in the high-frequency region compared with the theoretical values from the seismic model with modified opacity calculated for $X_{\rm {ini}}=0.70$ and $Z=0.013$. The rotational velocity at the stellar surface is $V_{\rm{rot}}=27.78$ km\,s$^{-1}$ (see Table\,\ref{tab:tOph:models}). We see that the frequencies of the quadrupole mode are pretty  well matched. In the case of the dipole mode, the theoretical frequencies are shifted to somewhat lower values  in comparison with the observed ones. This shift could be caused by higher order rotational effects, which were not included in our analysis.

\begin{figure}
	\includegraphics[width=\columnwidth]{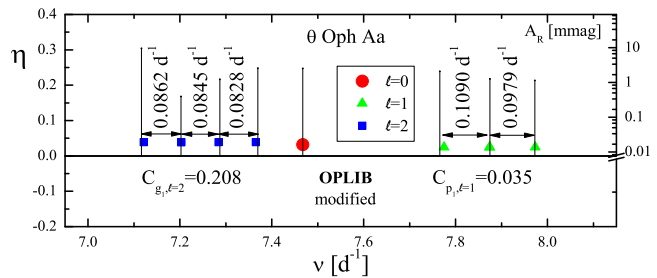}
	\caption{Instability parameter as a function of the frequency for the opacity modified model  calculated with $X_{\rm {ini}}=0.70$ and $Z=0.013$. Only p modes are shown. Separations of quadrupole and dipole multiplets are marked. Surface rotational velocity of the model is equal to $V_{\rm{rot}}=27.78$ kms$^{-1}$.}
	\label{fig:tOph_nu_eta_Cnl}
\end{figure}

The possibility of explaining the splittings in both multiplets by the same value of the rotational rate
suggests that the angular rotational velocity inside the radiative envelope is constant. This result is consistent with conclusions of \cite{2007MNRAS.381.1482B} and gives a strong constraint on the surface rotational velocity. The determined value of
the surface rotational velocity, $V_{\rm {rot}}\approx28$ km\,s$^{-1}$, is consistent within the error with the observed projected rotational velocity
$V_{\rm{rot}}\sin{i}\approx30\pm 9$ km\,s$^{-1}$, which corresponds to the minimum value of $V_{\rm{rot}}$. Taking the lower limit, i.e. $V_{\rm rot}=21$ km\,s$^{-1}$,
we get the allowed range of the inclination angle $i\in \langle 49^{\circ},90^{\circ}\rangle$. This range agrees with the value of $i$ derived from spectroscopic and astrometric orbits.
If the star is observed at $i=52^{\circ}$, then this may be an explanation for such a small amplitude of the frequency $\nu_{8}$,
because this value of $i$ corresponds to the node of the axisymmetric $\ell=2$ mode that is most probably associated with $\nu_{8}$.

As we have already mentioned  above, all our seismic models of $\theta$~Oph  Aa are located slightly outside the error box in the HR diagram (Fig.\,\ref{fig:HRtOph}). The empirical and model effective temperatures are marginally consistent (with the exception of the OP model) but the model luminosities are too large. The four seismic models of $\theta$~Oph Aa are marked in the HR diagram in Fig.\,\ref{fig:HRtOph} as circles. We chose the models with standard and modified opacity calculated with $X_{\rm {ini}}=0.70$ and $Z=0.013$ as well as the standard models calculated with the OP and OPAL opacities. Other seismic models occupy similar positions in the HR diagram. The position of the seismic models in the HR diagram is fixed mainly by the radial mode, $\nu_3$. Fitting its value provides a strong constraint on the mean stellar density. Models with a given value of the mean density are located along the line more or less parallel to the ZAMS  and their positions are, to a large extent, independent of the stellar parameters. The discrepancy is a strong indication that the Hipparcos parallax for $\theta$~Oph is overestimated. The newly released Gaia DR2 parallax \citep{2016A&A...595A...1G,
2018A&A...616A...1G,2018A&A...616A...2L} of $\theta$~Oph has, unfortunately, a large error and  is currently unable to resolve this issue.


\subsection{The $\theta$~Oph B hypothesis}
\label{sec:tOphB}

Although, the speckle  B5 companion contributes some 5-10\% to the total flux, it makes sense to consider that
the observed g modes originate in $\theta$~Oph B, especially in view of the fact that this star is inside the SPB instability strip \citep{1999AcA....49..119P,2017MNRAS.469...13S}.

We chose two models calculated with $X_{\rm {ini}}=0.70$, $Z=0.013$ and $M=5.4M_{\sun}$ that have the same age
as models of $\theta$~Oph Aa. The age of the first model amounts to  $\log{t/\rm{yr}}=7.305$ (corresponding to the standard model of  $\theta$~Oph Aa), the second one to $\log{t/\rm{yr}}=7.264$ (corresponding to the model with modified opacities). It turns out that g modes with frequencies in the observed range are stable in both non-rotating models of $\theta$~Oph B. The instability occurs for higher frequencies, i.e., for $\nu\gtrsim1$ d$^{-1}$. This can be seen in the bottom panel of Fig.\,\ref{fig:tOph_nu_eta_std_A_B}, where we plot the instability parameter for the older model of $\theta$~Oph B. The younger model does not differ significantly. The instability in the g-mode region is, however, much larger than in the more massive models of $\theta$~Oph Aa.

The rotational velocity of the second component is not constrained. Therefore, we considered a few values of  $V_{\rm {rot}}$
and computed pulsational models in the framework of the traditional approximation.
It turned out that even the standard opacity models calculated with the rotational velocity of $V_{\rm {rot}}=150$ km\,s$^{-1}$are able to account for the instability in the observed low-frequency region. The run of the instability parameter $\eta(\nu)$ can be seen in Fig.\,\ref{fig:tOphB_std_TA_150} in the g mode region. Modes with the harmonic degree $\ell=1$ and 2 were considered
together with all possible azimuthal orders.
The frequencies of very high-order $(\ell,m)$ = $(2,-2)$ modes exceeded the rotational frequency, which resulted in a reflection of  $\eta(\nu)$ at $\nu=0$.

 \begin{figure}
 	\includegraphics[width=\columnwidth]{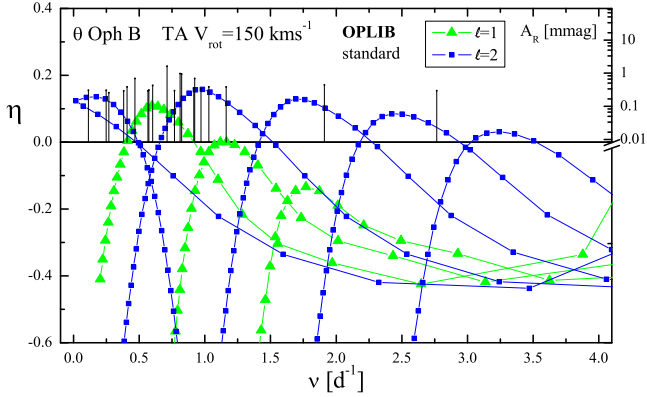}
 	 	\caption{The instability parameter, $\eta$,  as a function of the frequency for the standard opacity model of $\theta$~Oph B. Theoretical modes with $\ell=1$ and 2 were calculated in the framework of  the traditional approximation assuming a solid body rotation
	with $V_{\rm{rot}}=150$ km\,s$^{-1}$.}
 	\label{fig:tOphB_std_TA_150}
 \end{figure}

The lowest observed frequency can be explained by both dipole and quadrupole modes. Interestingly, the  frequency $\nu_{T}=2.76640$ d$^{-1}$, which was stable in the $\theta$~Oph Aa seismic  models, is excited in the model of $\theta$~Oph B. It is possible, therefore, that at least some low frequencies come from the secondary component of the $\theta$~Oph system. The advantage of this assumption is that the smaller mass models of $\theta$~Oph B do not need any modification of the standard opacity data. The necessary requirement is a rather fast rotational velocity of the order of 150 kms$^{-1}$. This hypothesis can be verified when disentangling the spectra and reliable determinations of $V_{\rm {rot}}$ for both components will be possible.

The models of $\theta$~Oph B are located marginally inside of the error box in the HR diagram. Both models are marked as  squares in Fig.\,\ref{fig:HRtOph}.

The sequence of frequencies quasi-equally spaced in period, which we described in Sect.\,\ref{sec:tOph_TA_gmodes}, cannot be associated with $\theta$~Oph B. Two of its components, $\nu_{M}$ and $\nu_{O}$,  originate in $\theta$~Oph Aa, as it has been noted in Sect.\,\ref{sec:fos}.

\section{Conclusions}
\label{sec:con}

From the analysis of space data of $\theta$~Oph from BRITE satellites we derived a lot of previously unknown frequencies. First of all, high-order g modes were discovered for the first time, which makes $\theta$~Oph another hybrid $\beta$~Cep/SPB pulsator with a rich frequency spectrum. In the p-mode region a very important frequency was detected corresponding to the centroid of the quadrupole mode. Thus, together with the radial mode and centroid of the dipole mode, we have three well-identified axisymmetric modes suitable for seismic modelling. Fitting these three frequencies gives strong constraints on stellar mass, effective temperature and overshooting from the convective core. These results depend on the metallicity, $Z$, which we chose as a free parameter. Adding the fourth frequency would fix the value of $Z$.

Because $\theta$~Oph is a triple system, with a low mass spectroscopic companion ($M_{Ab}<1M_{\sun}$}) and the speckle B5-type companion B, we considered two hypotheses.
Firstly, we assumed that all frequencies come from the primary B2-type component, $\theta$~Oph Aa. Assigning low-frequency g modes to $\theta$~Oph Aa imposes strong
constraints on the model parameters and opacities. This is because in all standard opacity models high-order g modes are stable.
We determined simultaneously corrections to the model and mean opacity profile near the $Z-$bump where driving of pulsations occurs.
To this end we searched for seismic models that 1) adjust the three centroid frequencies, 2) account for high-order g-mode instability and 3) reproduce the empirical value of the bolometric flux variations (the so-called parameter $f$).

For the  solar-like metallicity, $Z_{\sun}=0.013$, we got the overshooting parameter  $\alpha_{\rm {ov}}=0.3$  and the mass of about $8.4-8.6M_{\sun}$.
To account for the instability of g modes, a significant modification of $\kappa(T)$ was indispensable. The opacities had to be enhanced by 30\% at the depth of $\log{T/\rm{K}}=5.06$ (the Kurucz bump),
by 65\% at $\log{T/\rm{K}}=5.30$ (the maximum of the $Z-$bump) and by 145\% at $\log{T/\rm{K}}=5.47$ (the maximum contribution of nickel to the $Z-$bump). A huge increase at the `nickel' opacity was
forced  by the condition on high-order g mode instability. The shallowest increase at $\log{T/\rm{K}}=5.06$ was needed mainly to adjust the empirical and theoretical values of $f$.
Moreover, to account for instability of all low-frequency peaks the effects of rotation had to be properly included, which has been done in the framework of the traditional approximation.

Reproducing the splittings of the two multiplets constrained the surface rotational velocity at 28 km\,s$^{-1}$. Since both multiplets belong to the pressure dominated modes, the derived value of the rotation rate concerns the radiative envelope, where such modes propagate. Also, the splittings of the dipole and quadrupole modes are consistent, which implies a constant value
of the rotational velocity inside the stellar envelope. 

All our seismic models of $\theta$~Oph Aa (both with standard and modified opacities) are slightly above the error box in the HR  diagram. In the reasonable range  of stellar parameters it was impossible to shift the position of models into the  observationally determined ranges. There are two obvious solutions to that problem. Firstly, the identification of modes can be wrong, in particular the frequency $\nu_3$ may have been misclassified as a radial mode. We consider this solution rather unlikely,
because the photometric amplitude ratios and phase differences clearly indicate $\ell=0$ \citep{2005MNRAS.362..612H,2009MNRAS.398.1961D}. The other possibility is the wrong value of the parallax. The value given by  \cite{2007A&A...474..653V} amounts to 7.48 $\pm$ 0.17 mas, but the system is triple (and was not resolved by Hipparcos), which might have affected the result. We regard this as the most likely explanation of the discrepancy. 

In the second scenario, we considered the B5-type component as the origin  of the g mode pulsations. This hypothesis is validated by the fact that $\theta$~Oph B lies inside the SPB instability strip. In this case, to get instability in the g mode region one has to assume quite high rotation of $V_{\rm rot}>100$ km\,s$^{-1}$ and no opacity modification is needed.
However, it should be remembered that $\theta$~Oph B contributes only some 5-10\% to the total flux, which would mean that real amplitudes would have to be much larger than those observed.

The most interesting and meaningful result of our studies is that the opacity enhancements suitable for $\theta$~Oph Aa are similar to the modifications required for the other hybrid pulsators: $\nu$ Eri \citep{2017MNRAS.466.2284D}, $\alpha$ Lup \citep{2017sbcs.conf..173W}, 12 Lac \citep{2017EPJWC.15206005W}, $\gamma$ Peg \citep{2017EPJWC.15206005W} and $\kappa$\,Sco (Walczak et al., in preparation). The question that still remains is: what is the source of such a huge modification of the opacity data?
Recently, a significant increase of the opacity coefficient has been obtained in massive models by \cite{2018A&A...610L..15H} through including diffusion of heavy elements which caused the accumulation of iron and nickel near the driving zone. Thus, the effect of diffusion may be --- at least partly --- a solution of the g-mode instability problem in the hybrid $\beta$ Cep/SPB pulsators.
Such studies would be of high importance for both astrophysics and microphysics and this is the goal of our future project.

\section*{Acknowledgments} We thank M.-A. Dupret for his report which helped to significantly improve the paper. This work was financially supported by the Polish National Science Centre grants 2013/08/S/ST9/00583 and 2015/17/B/ST9/02082. APi acknowledges support from the NCN grant 2016/21/B/ST9/01126. AFJM is grateful for financial aid from NSERC (Canada) and FQRNT (Qu\'ebec). GAW acknowledges Discovery Grant support from the Natural Sciences and Engineering Research Council of Canada. GH acknowledges support from the Polish National Science Center (NCN), grant no. 2015/18/A/ST9/00578. Adam Popowicz was responsible for image processing and automatation of photometric routines for the data registered by BRITE-nanosatellite constellation, and was supported by NCN grant 2016/21/D/ST9/00656. K.Z. and W.W. acknowledges support by the Austrian Space Application Programme (ASAP) of the Austrian Research Promotion Agency (FFG). Calculations have been partly carried out using resources provided by Wroclaw Centre for Networking and Supercomputing (http://www.wcss.pl), grant no. 265. The operation of the Polish BRITE satellites is secured by a SPUB grant of MNiSW.




\bibliographystyle{mnras}
\bibliography{PWBiblio} 

\begin{thebibliography}{}
\makeatletter
\relax
\def\mn@urlcharsother{\let\do\@makeother \do\$\do\&\do\#\do\^\do\_\do\%\do\~}
\def\mn@doi{\begingroup\mn@urlcharsother \@ifnextchar [ {\mn@doi@}
  {\mn@doi@[]}}
\def\mn@doi@[#1]#2{\def\@tempa{#1}\ifx\@tempa\@empty \href
  {http://dx.doi.org/#2} {doi:#2}\else \href {http://dx.doi.org/#2} {#1}\fi
  \endgroup}
\def\mn@eprint#1#2{\mn@eprint@#1:#2::\@nil}
\def\mn@eprint@arXiv#1{\href {http://arxiv.org/abs/#1} {{\tt arXiv:#1}}}
\def\mn@eprint@dblp#1{\href {http://dblp.uni-trier.de/rec/bibtex/#1.xml}
  {dblp:#1}}
\def\mn@eprint@#1:#2:#3:#4\@nil{\def\@tempa {#1}\def\@tempb {#2}\def\@tempc
  {#3}\ifx \@tempc \@empty \let \@tempc \@tempb \let \@tempb \@tempa \fi \ifx
  \@tempb \@empty \def\@tempb {arXiv}\fi \@ifundefined
  {mn@eprint@\@tempb}{\@tempb:\@tempc}{\expandafter \expandafter \csname
  mn@eprint@\@tempb\endcsname \expandafter{\@tempc}}}

\bibitem[\protect\citeauthoryear{{Abt}, {Levato}  \& {Grosso}}{{Abt}
  et~al.}{2002}]{2002ApJ...573..359A}
{Abt} H.~A.,  {Levato} H.,   {Grosso} M.,  2002, \mn@doi [\apj]
  {10.1086/340590}, \href {http://adsabs.harvard.edu/abs/2002ApJ...573..359A}
  {573, 359}

\bibitem[\protect\citeauthoryear{{Asplund}, {Grevesse}, {Sauval}  \&
  {Scott}}{{Asplund} et~al.}{2009}]{AGSS09}
{Asplund} M.,  {Grevesse} N.,  {Sauval} A.~J.,   {Scott} P.,  2009, \mn@doi
  [ARAA] {10.1146/annurev.astro.46.060407.145222}, \href
  {http://adsabs.harvard.edu/abs/2009ARA%26A..47..481A} {47, 481}

\bibitem[\protect\citeauthoryear{{Baker} \& {Kippenhahn}}{{Baker} \&
  {Kippenhahn}}{1962}]{1962ZA.....54..114B}
{Baker} N.,  {Kippenhahn} R.,  1962, \zap, \href
  {http://adsabs.harvard.edu/abs/1962ZA.....54..114B} {54, 114}

\bibitem[\protect\citeauthoryear{{Balona}, {Daszy{\'n}ska-Daszkiewicz}  \&
  {Pamyatnykh}}{{Balona} et~al.}{2015}]{2015MNRAS.452.3073B}
{Balona} L.~A.,  {Daszy{\'n}ska-Daszkiewicz} J.,   {Pamyatnykh} A.~A.,  2015,
  \mn@doi [\mnras] {10.1093/mnras/stv1513}, \href
  {http://adsabs.harvard.edu/abs/2015MNRAS.452.3073B} {452, 3073}

\bibitem[\protect\citeauthoryear{{Bildsten}, {Ushomirsky}  \&
  {Cutler}}{{Bildsten} et~al.}{1996}]{1996ApJ...460..827B}
{Bildsten} L.,  {Ushomirsky} G.,   {Cutler} C.,  1996, \mn@doi [\apj]
  {10.1086/177012}, \href {http://adsabs.harvard.edu/abs/1996ApJ...460..827B}
  {460, 827}

\bibitem[\protect\citeauthoryear{{Briquet}, {Lefever}, {Uytterhoeven}  \&
  {Aerts}}{{Briquet} et~al.}{2005}]{2005MNRAS.362..619B}
{Briquet} M.,  {Lefever} K.,  {Uytterhoeven} K.,   {Aerts} C.,  2005, \mn@doi
  [\mnras] {10.1111/j.1365-2966.2005.09287.x}, \href
  {http://adsabs.harvard.edu/abs/2005MNRAS.362..619B} {362, 619}

\bibitem[\protect\citeauthoryear{{Briquet}, {Morel}, {Thoul}, {Scuflaire},
  {Miglio}, {Montalb{\'a}n}, {Dupret}  \& {Aerts}}{{Briquet}
  et~al.}{2007}]{2007MNRAS.381.1482B}
{Briquet} M.,  {Morel} T.,  {Thoul} A.,  {Scuflaire} R.,  {Miglio} A.,
  {Montalb{\'a}n} J.,  {Dupret} M.-A.,   {Aerts} C.,  2007, \mn@doi [\mnras]
  {10.1111/j.1365-2966.2007.12142.x}, \href
  {http://adsabs.harvard.edu/abs/2007MNRAS.381.1482B} {381, 1482}

\bibitem[\protect\citeauthoryear{{Brown} \& {Verschueren}}{{Brown} \&
  {Verschueren}}{1997}]{1997A&A...319..811B}
{Brown} A.~G.~A.,  {Verschueren} W.,  1997, \aap, \href
  {http://adsabs.harvard.edu/abs/1997A%26A...319..811B} {319, 811}

\bibitem[\protect\citeauthoryear{{Buldgen} et~al.,}{{Buldgen}
  et~al.}{2017}]{2017A&A...607A..58B}
{Buldgen} G.,  et~al., 2017, \mn@doi [\aap] {10.1051/0004-6361/201731354},
  \href {http://adsabs.harvard.edu/abs/2017A%26A...607A..58B} {607, A58}

\bibitem[\protect\citeauthoryear{{Chapman} \& {Lindzen}}{{Chapman} \&
  {Lindzen}}{1970}]{1970attg.book.....C}
{Chapman} S.,  {Lindzen} R.,  1970, {Atmospheric tides. Thermal and
  gravitational}.
Dordrecht: Reidel, 1970

\bibitem[\protect\citeauthoryear{{Charpinet}, {Fontaine}, {Brassard}  \&
  {Dorman}}{{Charpinet} et~al.}{1996}]{1996ApJ...471L.103C}
{Charpinet} S.,  {Fontaine} G.,  {Brassard} P.,   {Dorman} B.,  1996, \mn@doi
  [\apjl] {10.1086/310335}, \href
  {http://adsabs.harvard.edu/abs/1996ApJ...471L.103C} {471, L103}

\bibitem[\protect\citeauthoryear{{Charpinet}, {Fontaine}, {Brassard}, {Chayer},
  {Rogers}, {Iglesias}  \& {Dorman}}{{Charpinet}
  et~al.}{1997}]{1997ApJ...483L.123C}
{Charpinet} S.,  {Fontaine} G.,  {Brassard} P.,  {Chayer} P.,  {Rogers} F.~J.,
  {Iglesias} C.~A.,   {Dorman} B.,  1997, \mn@doi [\apjl] {10.1086/310741},
  \href {http://adsabs.harvard.edu/abs/1997ApJ...483L.123C} {483, L123}

\bibitem[\protect\citeauthoryear{{Colgan}, {Kilcrease}, {Magee}, {Abdallah},
  {Sherrill}, {Fontes}, {Hakel}  \& {Zhang}}{{Colgan} et~al.}{2015}]{OPLIB2}
{Colgan} J.,  {Kilcrease} D.~P.,  {Magee} N.~H.,  {Abdallah} J.,  {Sherrill}
  M.~E.,  {Fontes} C.~J.,  {Hakel} P.,   {Zhang} H.~L.,  2015, \mn@doi [High
  Energy Density Physics] {10.1016/j.hedp.2015.02.006}, \href
  {http://adsabs.harvard.edu/abs/2015HEDP...14...33C} {14, 33}

\bibitem[\protect\citeauthoryear{{Colgan} et~al.,}{{Colgan}
  et~al.}{2016}]{OPLIB1}
{Colgan} J.,  et~al., 2016, \mn@doi [\apj] {10.3847/0004-637X/817/2/116}, \href
  {http://adsabs.harvard.edu/abs/2016ApJ...817..116C} {817, 116}

\bibitem[\protect\citeauthoryear{{Cox}, {Morgan}, {Rogers}  \&
  {Iglesias}}{{Cox} et~al.}{1992}]{1992ApJ...393..272C}
{Cox} A.~N.,  {Morgan} S.~M.,  {Rogers} F.~J.,   {Iglesias} C.~A.,  1992,
  \mn@doi [\apj] {10.1086/171504}, \href
  {http://adsabs.harvard.edu/abs/1992ApJ...393..272C} {393, 272}

\bibitem[\protect\citeauthoryear{{Cugier}}{{Cugier}}{2012}]{2012A&A...547A..42C}
{Cugier} H.,  2012, \mn@doi [\aap] {10.1051/0004-6361/201219168}, \href
  {http://adsabs.harvard.edu/abs/2012A%26A...547A..42C} {547, A42}

\bibitem[\protect\citeauthoryear{{Cugier}}{{Cugier}}{2014}]{2014A&A...565A..76C}
{Cugier} H.,  2014, \mn@doi [\aap] {10.1051/0004-6361/201220507}, \href
  {http://adsabs.harvard.edu/abs/2014A%26A...565A..76C} {565, A76}

\bibitem[\protect\citeauthoryear{{Daszy{\'n}ska-Daszkiewicz} \&
  {Walczak}}{{Daszy{\'n}ska-Daszkiewicz} \&
  {Walczak}}{2009}]{2009MNRAS.398.1961D}
{Daszy{\'n}ska-Daszkiewicz} J.,  {Walczak} P.,  2009, \mn@doi [\mnras]
  {10.1111/j.1365-2966.2009.15229.x}, \href
  {http://adsabs.harvard.edu/abs/2009MNRAS.398.1961D} {398, 1961}

\bibitem[\protect\citeauthoryear{{Daszy{\'n}ska-Daszkiewicz}, {Dziembowski}  \&
  {Pamyatnykh}}{{Daszy{\'n}ska-Daszkiewicz} et~al.}{2003}]{JDD2003}
{Daszy{\'n}ska-Daszkiewicz} J.,  {Dziembowski} W.~A.,   {Pamyatnykh} A.~A.,
  2003, \mn@doi [A\&A] {10.1051/0004-6361:20030947}, \href
  {http://adsabs.harvard.edu/abs/2003A%26A...407..999D} {407, 999}

\bibitem[\protect\citeauthoryear{{Daszy{\'n}ska-Daszkiewicz}, {Dziembowski}  \&
  {Pamyatnykh}}{{Daszy{\'n}ska-Daszkiewicz} et~al.}{2005}]{JDD2005}
{Daszy{\'n}ska-Daszkiewicz} J.,  {Dziembowski} W.~A.,   {Pamyatnykh} A.~A.,
  2005, \mn@doi [A\&A] {10.1051/0004-6361:20052690}, \href
  {http://adsabs.harvard.edu/abs/2005A%26A...441..641D} {441, 641}

\bibitem[\protect\citeauthoryear{{Daszynska-Daszkiewicz}, {Dziembowski}  \&
  {Pamyatnykh}}{{Daszynska-Daszkiewicz} et~al.}{2007}]{2007AcA....57...11D}
{Daszynska-Daszkiewicz} J.,  {Dziembowski} W.~A.,   {Pamyatnykh} A.~A.,  2007,
  \actaa, \href {http://adsabs.harvard.edu/abs/2007AcA....57...11D} {57, 11}

\bibitem[\protect\citeauthoryear{{Daszy{\'n}ska-Daszkiewicz}, {Pamyatnykh},
  {Walczak}  \& {et al.}}{{Daszy{\'n}ska-Daszkiewicz}
  et~al.}{2017}]{2017MNRAS.466.2284D}
{Daszy{\'n}ska-Daszkiewicz} J.,  {Pamyatnykh} A.~A.,  {Walczak} P.,   {et al.}
  2017, \mn@doi [MNRAS] {10.1093/mnras/stw3315}, \href
  {http://adsabs.harvard.edu/abs/2017MNRAS.466.2284D} {466, 2284}

\bibitem[\protect\citeauthoryear{{Dziembowski}}{{Dziembowski}}{1977}]{1977AcA....27...95D}
{Dziembowski} W.,  1977, \actaa, \href
  {http://adsabs.harvard.edu/abs/1977AcA....27...95D} {27, 95}

\bibitem[\protect\citeauthoryear{{Dziembowski} \& {Pamiatnykh}}{{Dziembowski}
  \& {Pamiatnykh}}{1993}]{1993MNRAS.262..204D}
{Dziembowski} W.~A.,  {Pamiatnykh} A.~A.,  1993, \mn@doi [\mnras]
  {10.1093/mnras/262.1.204}, \href
  {http://adsabs.harvard.edu/abs/1993MNRAS.262..204D} {262, 204}

\bibitem[\protect\citeauthoryear{{Dziembowski} \& {Pamyatnykh}}{{Dziembowski}
  \& {Pamyatnykh}}{2008}]{2008MNRAS.385.2061D}
{Dziembowski} W.~A.,  {Pamyatnykh} A.~A.,  2008, \mn@doi [\mnras]
  {10.1111/j.1365-2966.2008.12964.x}, \href
  {http://adsabs.harvard.edu/abs/2008MNRAS.385.2061D} {385, 2061}

\bibitem[\protect\citeauthoryear{{Dziembowski}, {Moskalik}  \&
  {Pamyatnykh}}{{Dziembowski} et~al.}{1993}]{1993MNRAS.265..588D}
{Dziembowski} W.~A.,  {Moskalik} P.,   {Pamyatnykh} A.~A.,  1993, \mn@doi
  [\mnras] {10.1093/mnras/265.3.588}, \href
  {http://adsabs.harvard.edu/abs/1993MNRAS.265..588D} {265, 588}

\bibitem[\protect\citeauthoryear{{Dziembowski}, {Daszy{\'n}ska-Daszkiewicz}  \&
  {Pamyatnykh}}{{Dziembowski} et~al.}{2007}]{2007MNRAS.374..248D}
{Dziembowski} W.~A.,  {Daszy{\'n}ska-Daszkiewicz} J.,   {Pamyatnykh} A.~A.,
  2007, \mn@doi [\mnras] {10.1111/j.1365-2966.2006.11139.x}, \href
  {http://adsabs.harvard.edu/abs/2007MNRAS.374..248D} {374, 248}

\bibitem[\protect\citeauthoryear{{ESA}}{{ESA}}{1997}]{1997ESASP1200.....E}
{ESA} ed. 1997, {The HIPPARCOS and TYCHO catalogues. Astrometric and
  photometric star catalogues derived from the ESA HIPPARCOS Space Astrometry
  Mission}  ESA Special Publication Vol. 1200

\bibitem[\protect\citeauthoryear{{Eyles} et~al.,}{{Eyles}
  et~al.}{2003}]{2003SoPh..217..319E}
{Eyles} C.~J.,  et~al., 2003, \mn@doi [\solphys]
  {10.1023/B:SOLA.0000006903.75671.49}, \href
  {http://cdsads.u-strasbg.fr/abs/2003SoPh..217..319E} {217, 319}

\bibitem[\protect\citeauthoryear{{Flower}}{{Flower}}{1996}]{1996ApJ...469..355F}
{Flower} P.~J.,  1996, \mn@doi [\apj] {10.1086/177785}, \href
  {http://adsabs.harvard.edu/abs/1996ApJ...469..355F} {469, 355}

\bibitem[\protect\citeauthoryear{{Gaia Collaboration} et~al.,}{{Gaia
  Collaboration} et~al.}{2016}]{2016A&A...595A...1G}
{Gaia Collaboration} et~al., 2016, \mn@doi [\aap]
  {10.1051/0004-6361/201629272}, \href
  {http://adsabs.harvard.edu/abs/2016A%26A...595A...1G} {595, A1}

\bibitem[\protect\citeauthoryear{{Gaia Collaboration} et~al.,}{{Gaia
  Collaboration} et~al.}{2018}]{2018A&A...616A...1G}
{Gaia Collaboration} et~al., 2018, \mn@doi [\aap]
  {10.1051/0004-6361/201833051}, \href
  {http://adsabs.harvard.edu/abs/2018A%26A...616A...1G} {616, A1}

\bibitem[\protect\citeauthoryear{{Handler}, {Shobbrook}  \&
  {Mokgwetsi}}{{Handler} et~al.}{2005}]{2005MNRAS.362..612H}
{Handler} G.,  {Shobbrook} R.~R.,   {Mokgwetsi} T.,  2005, \mn@doi [\mnras]
  {10.1111/j.1365-2966.2005.09341.x}, \href
  {http://adsabs.harvard.edu/abs/2005MNRAS.362..612H} {362, 612}

\bibitem[\protect\citeauthoryear{{Handler} et~al.,}{{Handler}
  et~al.}{2017}]{2017MNRAS.464.2249H}
{Handler} G.,  et~al., 2017, \mn@doi [\mnras] {10.1093/mnras/stw2518}, \href
  {http://adsabs.harvard.edu/abs/2017MNRAS.464.2249H} {464, 2249}

\bibitem[\protect\citeauthoryear{{Henroteau}}{{Henroteau}}{1922}]{1922PDO.....8....1H}
{Henroteau} F.,  1922, Publications of the Dominion Observatory Ottawa, \href
  {http://adsabs.harvard.edu/abs/1922PDO.....8....1H} {8, 1}

\bibitem[\protect\citeauthoryear{{Hick}, {Buffington}  \& {Jackson}}{{Hick}
  et~al.}{2007}]{2007SPIE.6689E..0CH}
{Hick} P.,  {Buffington} A.,   {Jackson} B.~V.,  2007, in Solar Physics and
  Space Weather Instrumentation II. p. 66890C, \mn@doi{10.1117/12.734808}

\bibitem[\protect\citeauthoryear{{Horch}, {van Belle}, {Davidson}, {Ciastko},
  {Everett}  \& {Bjorkman}}{{Horch} et~al.}{2015}]{2015AJ....150..151H}
{Horch} E.~P.,  {van Belle} G.~T.,  {Davidson} Jr. J.~W.,  {Ciastko} L.~A.,
  {Everett} M.~E.,   {Bjorkman} K.~S.,  2015, \mn@doi [\aj]
  {10.1088/0004-6256/150/5/151}, \href
  {http://cdsads.u-strasbg.fr/abs/2015AJ....150..151H} {150, 151}

\bibitem[\protect\citeauthoryear{{Hui-Bon-Hoa} \& {Vauclair}}{{Hui-Bon-Hoa} \&
  {Vauclair}}{2018}]{2018A&A...610L..15H}
{Hui-Bon-Hoa} A.,  {Vauclair} S.,  2018, \mn@doi [\aap]
  {10.1051/0004-6361/201832706}, \href
  {http://adsabs.harvard.edu/abs/2018A%26A...610L..15H} {610, L15}

\bibitem[\protect\citeauthoryear{{Iglesias} \& {Rogers}}{{Iglesias} \&
  {Rogers}}{1991}]{1991ApJ...371L..73I}
{Iglesias} C.~A.,  {Rogers} F.~J.,  1991, \mn@doi [\apjl] {10.1086/186005},
  \href {http://adsabs.harvard.edu/abs/1991ApJ...371L..73I} {371, L73}

\bibitem[\protect\citeauthoryear{{Iglesias} \& {Rogers}}{{Iglesias} \&
  {Rogers}}{1996}]{OPAL}
{Iglesias} C.~A.,  {Rogers} F.~J.,  1996, \mn@doi [\apj] {10.1086/177381},
  \href {http://adsabs.harvard.edu/abs/1996ApJ...464..943I} {464, 943}

\bibitem[\protect\citeauthoryear{{Irwin}}{{Irwin}}{1952}]{1952ApJ...116..211I}
{Irwin} J.~B.,  1952, \mn@doi [\apj] {10.1086/145604}, \href
  {http://cdsads.u-strasbg.fr/abs/1952ApJ...116..211I} {116, 211}

\bibitem[\protect\citeauthoryear{{Jackson} et~al.,}{{Jackson}
  et~al.}{2004}]{2004SoPh..225..177J}
{Jackson} B.~V.,  et~al., 2004, \mn@doi [\solphys] {10.1007/s11207-004-2766-3},
  \href {http://cdsads.u-strasbg.fr/abs/2004SoPh..225..177J} {225, 177}

\bibitem[\protect\citeauthoryear{{Kallinger} et~al.,}{{Kallinger}
  et~al.}{2017}]{2017A&A...603A..13K}
{Kallinger} T.,  et~al., 2017, \mn@doi [\aap] {10.1051/0004-6361/201730625},
  \href {http://cdsads.u-strasbg.fr/abs/2017A%26A...603A..13K} {603, A13}

\bibitem[\protect\citeauthoryear{{Kiriakidis}, {El Eid}  \&
  {Glatzel}}{{Kiriakidis} et~al.}{1992}]{1992MNRAS.255P...1K}
{Kiriakidis} M.,  {El Eid} M.~F.,   {Glatzel} W.,  1992, \mn@doi [\mnras]
  {10.1093/mnras/255.1.1P}, \href
  {http://adsabs.harvard.edu/abs/1992MNRAS.255P...1K} {255, 1P}

\bibitem[\protect\citeauthoryear{{Kurucz}}{{Kurucz}}{2004}]{Kurucz2004}
{Kurucz} R.,  2004, http:// kurucz.harvard.edu

\bibitem[\protect\citeauthoryear{{Kurucz}}{{Kurucz}}{2005}]{2005MSAIS...8...14K}
{Kurucz} R.~L.,  2005, Memorie della Societa Astronomica Italiana Supplementi,
  \href {http://adsabs.harvard.edu/abs/2005MSAIS...8...14K} {8, 14}

\bibitem[\protect\citeauthoryear{{Kurucz}}{{Kurucz}}{2011}]{2011CaJPh..89..417K}
{Kurucz} R.~L.,  2011, \mn@doi [Canadian Journal of Physics] {10.1139/p10-104},
  \href {http://adsabs.harvard.edu/abs/2011CaJPh..89..417K} {89, 417}

\bibitem[\protect\citeauthoryear{{Lanz} \& {Hubeny}}{{Lanz} \&
  {Hubeny}}{2007}]{2007ApJS..169...83L}
{Lanz} T.,  {Hubeny} I.,  2007, \mn@doi [\apj] {10.1086/511270}, \href
  {http://adsabs.harvard.edu/abs/2007ApJS..169...83L} {169, 83}

\bibitem[\protect\citeauthoryear{{Ledoux}}{{Ledoux}}{1951}]{1951ApJ...114..373L}
{Ledoux} P.,  1951, \mn@doi [\apj] {10.1086/145477}, \href
  {http://adsabs.harvard.edu/abs/1951ApJ...114..373L} {114, 373}

\bibitem[\protect\citeauthoryear{{Lee} \& {Saio}}{{Lee} \&
  {Saio}}{1997}]{1997ApJ...491..839L}
{Lee} U.,  {Saio} H.,  1997, \mn@doi [\apj] {10.1086/304980}, \href
  {http://adsabs.harvard.edu/abs/1997ApJ...491..839L} {491, 839}

\bibitem[\protect\citeauthoryear{{Lindegren} et~al.,}{{Lindegren}
  et~al.}{2018}]{2018A&A...616A...2L}
{Lindegren} L.,  et~al., 2018, \mn@doi [\aap] {10.1051/0004-6361/201832727},
  \href {http://adsabs.harvard.edu/abs/2018A%26A...616A...2L} {616, A2}

\bibitem[\protect\citeauthoryear{{Loader} et~al.,}{{Loader}
  et~al.}{2012}]{2012JDSO...8..313L}
{Loader} B.,  et~al., 2012, Journal of Double Star Observations, 8, 313

\bibitem[\protect\citeauthoryear{{Lovekin} \& {Goupil}}{{Lovekin} \&
  {Goupil}}{2010}]{2010A&A...515A..58L}
{Lovekin} C.~C.,  {Goupil} M.-J.,  2010, \mn@doi [\aap]
  {10.1051/0004-6361/200913855}, \href
  {http://adsabs.harvard.edu/abs/2010A%26A...515A..58L} {515, A58}

\bibitem[\protect\citeauthoryear{{McAlister}, {Mason}, {Hartkopf}  \&
  {Shara}}{{McAlister} et~al.}{1993}]{1993AJ....106.1639M}
{McAlister} H.~A.,  {Mason} B.~D.,  {Hartkopf} W.~I.,   {Shara} M.~M.,  1993,
  \mn@doi [\aj] {10.1086/116753}, \href
  {http://adsabs.harvard.edu/abs/1993AJ....106.1639M} {106, 1639}

\bibitem[\protect\citeauthoryear{{McNamara}}{{McNamara}}{1957}]{1957PASP...69..570M}
{McNamara} D.~H.,  1957, \mn@doi [\pasp] {10.1086/127152}, \href
  {http://adsabs.harvard.edu/abs/1957PASP...69..570M} {69, 570}

\bibitem[\protect\citeauthoryear{{Moskalik} \& {Dziembowski}}{{Moskalik} \&
  {Dziembowski}}{1992}]{1992A&A...256L...5M}
{Moskalik} P.,  {Dziembowski} W.~A.,  1992, \aap, \href
  {http://adsabs.harvard.edu/abs/1992A%26A...256L...5M} {256, L5}

\bibitem[\protect\citeauthoryear{{Niemczura} \&
  {Daszy{\'n}ska-Daszkiewicz}}{{Niemczura} \&
  {Daszy{\'n}ska-Daszkiewicz}}{2005}]{2005A&A...433..659N}
{Niemczura} E.,  {Daszy{\'n}ska-Daszkiewicz} J.,  2005, \mn@doi [\aap]
  {10.1051/0004-6361:20040396}, \href
  {http://adsabs.harvard.edu/abs/2005A%26A...433..659N} {433, 659}

\bibitem[\protect\citeauthoryear{{Nieva} \& {Przybilla}}{{Nieva} \&
  {Przybilla}}{2012}]{2012A&A...539A.143N}
{Nieva} M.-F.,  {Przybilla} N.,  2012, \mn@doi [\aap]
  {10.1051/0004-6361/201118158}, \href
  {http://adsabs.harvard.edu/abs/2012A%26A...539A.143N} {539, A143}

\bibitem[\protect\citeauthoryear{{Pablo} et~al.,}{{Pablo}
  et~al.}{2016}]{2016PASP..128l5001P}
{Pablo} H.,  et~al., 2016, \mn@doi [\pasp] {10.1088/1538-3873/128/970/125001},
  \href {http://adsabs.harvard.edu/abs/2016PASP..128l5001P} {128, 125001}

\bibitem[\protect\citeauthoryear{{Pamyatnykh}}{{Pamyatnykh}}{1999}]{1999AcA....49..119P}
{Pamyatnykh} A.~A.,  1999, \actaa, \href
  {http://adsabs.harvard.edu/abs/1999AcA....49..119P} {49, 119}

\bibitem[\protect\citeauthoryear{{Pamyatnykh}, {Dziembowski}, {Handler}  \&
  {Pikall}}{{Pamyatnykh} et~al.}{1998}]{1998A&A...333..141P}
{Pamyatnykh} A.~A.,  {Dziembowski} W.~A.,  {Handler} G.,   {Pikall} H.,  1998,
  \aap, \href {http://adsabs.harvard.edu/abs/1998A%26A...333..141P} {333, 141}

\bibitem[\protect\citeauthoryear{{Pamyatnykh}, {Handler}  \&
  {Dziembowski}}{{Pamyatnykh} et~al.}{2004}]{2004MNRAS.350.1022P}
{Pamyatnykh} A.~A.,  {Handler} G.,   {Dziembowski} W.~A.,  2004, \mn@doi
  [MNRAS] {10.1111/j.1365-2966.2004.07721.x}, \href
  {http://adsabs.harvard.edu/abs/2004MNRAS.350.1022P} {350, 1022}

\bibitem[\protect\citeauthoryear{{Pigulski} \& {the BRITE Team}}{{Pigulski} \&
  {the BRITE Team}}{2018}]{2018adlc175}
{Pigulski} A.,  {the BRITE Team} 2018, in {Wade} G.~A.,  {Baade} D.,  {Guzik}
  J.~A.,   {Smolec} R.,  eds,  Proceedings of the Polish Astronomical Society
  Vol. 7, Proceedings of the Polish Astronomical Society. pp 175--190
  (\mn@eprint {arXiv} {1801.08496})

\bibitem[\protect\citeauthoryear{{Pigulski} et~al.,}{{Pigulski}
  et~al.}{2016}]{2016A&A...588A..55P}
{Pigulski} A.,  et~al., 2016, \mn@doi [\aap] {10.1051/0004-6361/201527872},
  \href {http://adsabs.harvard.edu/abs/2016A%26A...588A..55P} {588, A55}

\bibitem[\protect\citeauthoryear{{Pigulski}, {Popowicz}, {Kuschnig}  \& {the
  BRITE Team}}{{Pigulski} et~al.}{2018}]{2018adlc106}
{Pigulski} A.,  {Popowicz} A.,  {Kuschnig} R.,   {the BRITE Team} 2018, in
  {Wade} G.~A.,  {Baade} D.,  {Guzik} J.~A.,   {Smolec} R.,  eds,  Proceedings
  of the Polish Astronomical Society Vol. 7, Proceedings of the Polish
  Astronomical Society. pp 106--114 (\mn@eprint {arXiv} {1802.09021})

\bibitem[\protect\citeauthoryear{{Popowicz} et~al.,}{{Popowicz}
  et~al.}{2017}]{2017A&A...605A..26P}
{Popowicz} A.,  et~al., 2017, \mn@doi [\aap] {10.1051/0004-6361/201730806},
  \href {http://adsabs.harvard.edu/abs/2017A%26A...605A..26P} {605, A26}

\bibitem[\protect\citeauthoryear{{Roberts} \& {Mason}}{{Roberts} \&
  {Mason}}{2018}]{2018MNRAS.473.4497R}
{Roberts} L.~C.,  {Mason} B.~D.,  2018, \mn@doi [\mnras]
  {10.1093/mnras/stx2559}, \href
  {http://cdsads.u-strasbg.fr/abs/2018MNRAS.473.4497R} {473, 4497}

\bibitem[\protect\citeauthoryear{{Rogers} \& {Nayfonov}}{{Rogers} \&
  {Nayfonov}}{2002}]{2002ApJ...576.1064R}
{Rogers} F.~J.,  {Nayfonov} A.,  2002, \mn@doi [\apj] {10.1086/341894}, \href
  {http://adsabs.harvard.edu/abs/2002ApJ...576.1064R} {576, 1064}

\bibitem[\protect\citeauthoryear{{Salmon}, {Montalb{\'a}n}, {Morel}  \& {et
  al.}}{{Salmon} et~al.}{2012}]{2012MNRAS.422.3460S}
{Salmon} S.,  {Montalb{\'a}n} J.,  {Morel} T.,   {et al.} 2012, \mn@doi [MNRAS]
  {10.1111/j.1365-2966.2012.20857.x}, \href
  {http://adsabs.harvard.edu/abs/2012MNRAS.422.3460S} {422, 3460}

\bibitem[\protect\citeauthoryear{{Seaton}}{{Seaton}}{2005}]{OP}
{Seaton} M.~J.,  2005, \mn@doi [MNRAS] {10.1111/j.1365-2966.2005.00019.x},
  \href {http://adsabs.harvard.edu/abs/2005MNRAS.362L...1S} {362, L1}

\bibitem[\protect\citeauthoryear{{Seaton}, {Zeippen}, {Tully}, {Pradhan},
  {Mendoza}, {Hibbert}  \& {Berrington}}{{Seaton}
  et~al.}{1992}]{1992RMxAA..23...19S}
{Seaton} M.~J.,  {Zeippen} C.~J.,  {Tully} J.~A.,  {Pradhan} A.~K.,  {Mendoza}
  C.,  {Hibbert} A.,   {Berrington} K.~A.,  1992, \rmxaa, \href
  {http://adsabs.harvard.edu/abs/1992RMxAA..23...19S} {23}

\bibitem[\protect\citeauthoryear{{Shatsky} \& {Tokovinin}}{{Shatsky} \&
  {Tokovinin}}{2002}]{2002A+A...382...92S}
{Shatsky} N.,  {Tokovinin} A.,  2002, \mn@doi [\aap]
  {10.1051/0004-6361:20011542}, \href
  {http://adsabs.harvard.edu/abs/2002A%26A...382...92S} {382, 92}

\bibitem[\protect\citeauthoryear{{Stankov} \& {Handler}}{{Stankov} \&
  {Handler}}{2005}]{2005ApJS..158..193S}
{Stankov} A.,  {Handler} G.,  2005, \mn@doi [\apjs] {10.1086/429408}, \href
  {http://adsabs.harvard.edu/abs/2005ApJS..158..193S} {158, 193}

\bibitem[\protect\citeauthoryear{{Stellingwerf}}{{Stellingwerf}}{1978}]{1978AJ.....83.1184S}
{Stellingwerf} R.~F.,  1978, \mn@doi [AJ] {10.1086/112308}, \href
  {http://adsabs.harvard.edu/abs/1978AJ.....83.1184S} {83, 1184}

\bibitem[\protect\citeauthoryear{{Szewczuk} \&
  {Daszy{\'n}ska-Daszkiewicz}}{{Szewczuk} \&
  {Daszy{\'n}ska-Daszkiewicz}}{2017}]{2017MNRAS.469...13S}
{Szewczuk} W.,  {Daszy{\'n}ska-Daszkiewicz} J.,  2017, \mn@doi [\mnras]
  {10.1093/mnras/stx738}, \href
  {http://adsabs.harvard.edu/abs/2017MNRAS.469...13S} {469, 13}

\bibitem[\protect\citeauthoryear{{Tokovinin}, {Mason}, {Hartkopf}, {Mendez}  \&
  {Horch}}{{Tokovinin} et~al.}{2015}]{2015AJ....150...50T}
{Tokovinin} A.,  {Mason} B.~D.,  {Hartkopf} W.~I.,  {Mendez} R.~A.,   {Horch}
  E.~P.,  2015, \mn@doi [\aj] {10.1088/0004-6256/150/2/50}, \href
  {http://cdsads.u-strasbg.fr/abs/2015AJ....150...50T} {150, 50}

\bibitem[\protect\citeauthoryear{{Townsend}}{{Townsend}}{2003a}]{2003MNRAS.340.1020T}
{Townsend} R.~H.~D.,  2003a, \mn@doi [\mnras]
  {10.1046/j.1365-8711.2003.06379.x}, \href
  {http://adsabs.harvard.edu/abs/2003MNRAS.340.1020T} {340, 1020}

\bibitem[\protect\citeauthoryear{{Townsend}}{{Townsend}}{2003b}]{2003MNRAS.343..125T}
{Townsend} R.~H.~D.,  2003b, \mn@doi [\mnras]
  {10.1046/j.1365-8711.2003.06640.x}, \href
  {http://adsabs.harvard.edu/abs/2003MNRAS.343..125T} {343, 125}

\bibitem[\protect\citeauthoryear{{Unno}, {Osaki}, {Ando}, {Saio}  \&
  {Shibahashi}}{{Unno} et~al.}{1989}]{1989nos..book.....U}
{Unno} W.,  {Osaki} Y.,  {Ando} H.,  {Saio} H.,   {Shibahashi} H.,  1989,
  {Nonradial oscillations of stars}.
University of Tokyo Press

\bibitem[\protect\citeauthoryear{{Walczak} \&
  {Daszy{\'n}ska-Daszkiewicz}}{{Walczak} \&
  {Daszy{\'n}ska-Daszkiewicz}}{2014}]{2014IAUS..301..221W}
{Walczak} P.,  {Daszy{\'n}ska-Daszkiewicz} J.,  2014, in {Guzik} J.~A.,
  {Chaplin} W.~J.,  {Handler} G.,   {Pigulski} A.,  eds,  IAU Symposium Vol.
  301, Precision Asteroseismology. pp 221--228 (\mn@eprint {arXiv}
  {1312.7668}), \mn@doi{10.1017/S1743921313014361}

\bibitem[\protect\citeauthoryear{{Walczak}, {Daszy{\'n}ska-Daszkiewicz},
  {Pamyatnykh}, {Handler}  \& {Pigulski}}{{Walczak}
  et~al.}{2017a}]{2017sbcs.conf..173W}
{Walczak} P.,  {Daszy{\'n}ska-Daszkiewicz} J.,  {Pamyatnykh} A.,  {Handler} G.,
    {Pigulski} A.,  2017a, in Second BRITE-Constellation Science Conference:
  Small satellites{\textemdash}big science, Proceedings of the Polish
  Astronomical Society volume 5, held 22-26 August, 2016 in Innsbruck, Austria.
  Other: Polish Astronomical Society, Bartycka 18, 00-716 Warsaw, Poland,
  pp.173-179. pp 173--179 (\mn@eprint {arXiv} {1701.01258})

\bibitem[\protect\citeauthoryear{{Walczak}, {Daszy{\'n}ska-Daszkiewicz}  \&
  {Pamyatnykh}}{{Walczak} et~al.}{2017b}]{2017EPJWC.15206005W}
{Walczak} P.,  {Daszy{\'n}ska-Daszkiewicz} J.,   {Pamyatnykh} A.,  2017b, in
  European Physical Journal Web of Conferences. p. 06005 (\mn@eprint {arXiv}
  {1704.06067}), \mn@doi{10.1051/epjconf/201715206005}

\bibitem[\protect\citeauthoryear{{Weiss} et~al.,}{{Weiss}
  et~al.}{2014}]{BRITE1}
{Weiss} W.~W.,  et~al., 2014, \mn@doi [\pasp] {10.1086/677236}, \href
  {http://adsabs.harvard.edu/abs/2014PASP..126..573W} {126, 573}

\bibitem[\protect\citeauthoryear{{Zhevakin}}{{Zhevakin}}{1963}]{1963ARA&A...1..367Z}
{Zhevakin} S.~A.,  1963, \mn@doi [\araa] {10.1146/annurev.aa.01.090163.002055},
  \href {http://adsabs.harvard.edu/abs/1963ARA%26A...1..367Z} {1, 367}

\bibitem[\protect\citeauthoryear{{van Hoof} \& {Blaauw}}{{van Hoof} \&
  {Blaauw}}{1958}]{1958ApJ...128..273V}
{van Hoof} A.,  {Blaauw} A.,  1958, \mn@doi [\apj] {10.1086/146542}, \href
  {http://adsabs.harvard.edu/abs/1958ApJ...128..273V} {128, 273}

\bibitem[\protect\citeauthoryear{{van Hoof}, {Bertiau}  \& {Deurinck}}{{van
  Hoof} et~al.}{1956}]{1956ApJ...124..168V}
{van Hoof} A.,  {Bertiau} F.,   {Deurinck} R.,  1956, \mn@doi [\apj]
  {10.1086/146212}, \href {http://adsabs.harvard.edu/abs/1956ApJ...124..168V}
  {124, 168}

\bibitem[\protect\citeauthoryear{{van Leeuwen}}{{van
  Leeuwen}}{2007}]{2007A&A...474..653V}
{van Leeuwen} F.,  2007, \mn@doi [\aap] {10.1051/0004-6361:20078357}, \href
  {http://adsabs.harvard.edu/abs/2007A%26A...474..653V} {474, 653}

\makeatother
\end{thebibliography}








\bsp	
\label{lastpage}
\end{document}